%% file: main.tex
\documentclass[11pt,a4paper]{article}
\usepackage[T1]{fontenc}
\usepackage[utf8]{inputenc}
\usepackage{lmodern}
\usepackage[margin=2.3cm]{geometry}
\usepackage{amsmath,amssymb,amsthm}
\usepackage{booktabs,array,longtable}
\usepackage{graphicx}
\usepackage{pifont}
\usepackage{xcolor}
\usepackage{enumitem}
\usepackage[hidelinks]{hyperref}
\usepackage{caption}
\newcommand{\cmark}{\textcolor[HTML]{1b7a3a}{\ding{51}}}
\newcommand{\xmark}{\textcolor[HTML]{b3261e}{\ding{55}}}
\newcommand{\pmark}{\textcolor[HTML]{8a6d00}{$\boldsymbol{\sim}$}}
\newcommand{\nf}{\mathrm{nf}}
\newcommand{\Ep}{\ensuremath{\mathrm{E}'}}
\newcommand{\Epl}{\ensuremath{\mathrm{E}^{+}}}
\newcommand{\metric}[2]{\textsf{#1}~(\emph{#2})}
\newcommand{\new}{$^{\star}$}

\theoremstyle{definition}

\title{How Far Apart Are Two Proofs?\\[2pt]
\large Measuring Proof Dissimilarity in a Proof Kernel and in Lean/Mathlib}
\author{Nikolaos Nikolaou\\\small University College London}
\date{23 November 2026}

\begin{document}
\maketitle

\begin{abstract}
\noindent
We now live in the age of machine-generated mathematics. AI systems produce research-level proofs at a scale and speed that the human mathematical community can no longer catch up with. Verifying them is only part of the challenge. We also need to know whether a ``new proof'' is indeed new, whether a thousand generated proofs contain one idea or many, and where each sits relative to human mathematics. All of these questions require \emph{measuring distances between proofs}. Hilbert's unpublished 24th problem and the identity-of-proofs problem of general proof theory ask when two proofs are ``the same''. In this work, we propose turning this into a graded question: how far apart are two proofs of one theorem, and what is the shape of the \emph{proof space} they inhabit? We define 19 dissimilarity measures in 6 families (syntactic, structural, dependency \& resource, semantic, information-theoretic, higher-categorical), 12 of them new or adapted.
We also define 10 desirable properties, which we argue are exactly the conditions under which a measure can be safely optimised. In a Curry--Howard kernel, 5 machine-checked proofs that the primes are infinite show three things. First, that raw syntax cannot tell a rewrite from a new argument. Second, that dependency-based and structural measures reproduce the intuitive ordering of distances from Euclid's proof (Euclid rewritten $<$ Goldbach $<$ Furstenberg). Third, that semantic measures invert it: run against a putative finite list of primes, Furstenberg's topological proof produces exactly Euclid's witness, the least prime factor of $1+\prod S$. A replication in Lean~4/Mathlib (3 theorems, 19 proofs, 18 automatically extracted measures) confirms that dependency-closure measures track intuition best (full ordering on all theorems, 9/10 properties). It also shows that equality of normal forms---the classical criterion of identity---never holds between differently written versions of the same argument, that proofs by citation look unlike the arguments they cite and that a proof assistant's own notion of identity (proof irrelevance) cannot distinguish any two proofs. Most measures also apply to proofs of \emph{different} theorems, where they currently capture shared style rather than shared ideas, and we propose anti-unification for detecting the same method across theorems. We argue that dissimilarity is best reported as a profile across levels of identification, and we discuss uses:
\begin{itemize}[leftmargin=*,itemsep=0pt,topsep=2pt]
\item detecting recalled or duplicated AI proofs;
\item orchestrating prover agents for complementarity;
\item finding the most similar human-understandable proof to interpret an AI proof;
\item mapping and exploring proof space.
\end{itemize}
All code, Lean sources and data are released.
\end{abstract}

\tableofcontents

\section{Introduction}\label{sec:intro}

\paragraph{Why the question has become urgent.}
Until recently, the alternative proofs of a theorem accumulated slowly, through the work of a few mathematicians over decades.
In 2026 this changed.
In August, Alp\"oge and Furman published a claimed proof, with a Lean formalisation, that more than $67.25\%$ of the non-trivial zeros of the Riemann zeta function are simple and lie on the critical line. The argument was discovered by an unreleased research version of Anthropic's Claude~\cite{wiki-rh}.
On 8~September, OpenAI announced that a swarm of about $10^4$ agents, running an unreleased internal model, had found in 88 hours a finite-time, finite-energy singularity of the forced three-dimensional Navier--Stokes equations from smooth data, the ``blow-up'' alternative of the Clay problem. It also announced a disproof of regularity for the unforced Euler equations. A Lean formalisation took a further 17 hours~\cite{oai-ns,quanta-ns}.
The announcement came shortly after Buckmaster and Alp\"oge, working with several AI models, had obtained a Lean-verified singularity result for the forced Euler equations. A public dispute followed over priority, over whether one effort built on the other, and over whether a user's interactions with an AI product could have informed a competitor's model~\cite{quanta-ns,techcrunch-ns,wiki-ns}.
In October, OpenAI released a repository of 721 machine-generated mathematical manuscripts, grouped into 373 \emph{families}, about 42\% of whose top-line results have Lean formalisations. The repository itself describes them as raw research data that may contain errors~\cite{oai-math}.
On 26~October it announced proofs of Sogge's local smoothing conjecture in all dimensions, and with it the Kakeya and restriction conjectures; a Lean-verified disproof of the exponential time hypothesis; and existence proofs for interacting four-dimensional gauge theories, including pure Yang--Mills (existence only, not the mass gap, Lean-formalised) and QCD. The QCD proofs are being written up by a group led by David Gross~\cite{oai-oct}.
Independent verification of most of these claims was still pending at the time of writing~\cite{wiki-ns,wiki-ym}.\footnote{All statements in this paragraph are taken from the cited announcements and reports, accessed 23~November~2026. Not all of these claims are proven---or rather verified---results yet.}

These events raise questions that a proof assistant's kernel cannot settle.\footnote{The kernel is the small, trusted core of a system such as Lean that checks every step of a formal proof, and it only certifies that a proof is correct.}
When two teams, human, machine or mixed, prove the same or closely related statements within weeks, is it the \emph{same proof}?
When a model trained partly on user interactions produces an argument close to one that a user was developing, how close is ``close''?
When ten thousand agents explore a problem in parallel, how many \emph{different} arguments did they find, and which partial results should be merged?
When 373 families of results appear at once, how do we place them relative to the existing literature, and to each other?
When an AI system finds a proof, how much of it was already present in the human-provided problem formulation, hints and sketches?
And when experts call a first machine-generated proof ``horrendous'' to read~\cite{quanta-ns}, can its distance from human-style arguments be measured? Could we perhaps make it more humanly understandable if we related it to its closest human-style arguments?
Perhaps more ambitiously, can we learn from similarities of patterns of strategies applied across different mathematical objects to fill in gaps in our mathematical knowledge? Ultimately all these are questions concerning the \emph{(dis)similarity of proofs}. Hilbert's 24th problem has turned from a philosophical curiosity into a practical need, and Section~\ref{sec:enable} develops the connection.

\paragraph{The identity of proofs.}

A theorem can have many proofs; Lemmermeyer catalogues more than two hundred for quadratic reciprocity.
Mathematicians judge, quickly and fairly consistently, that some pairs of proofs are ``really the same'', some are ``variations on a theme'', and some ``have nothing in common''.
Making this judgement precise is old.
Hilbert drafted a 24th problem, asking for criteria of simplicity and of sameness of proofs, but left it out of his 1900 list~\cite{thiele}.
General proof theory took the question up as the \emph{identity of proofs} problem~\cite{prawitz,kreisel,dosen}.
Proposed answers include equality of normal forms under Curry--Howard, equality of morphisms in free categories~\cite{lambek}, equality of proof nets~\cite{girard,strassburger}, and equality of paths in homotopy type theory~\cite{hott}.

These proposals give an \emph{equivalence relation}, a yes/no answer.
They say nothing about \emph{how far apart} two non-identified proofs are.
We could not find a systematic comparison of candidate \emph{distances} between proofs.
This work provides one.
\begin{enumerate}[leftmargin=*,itemsep=1pt]
\item We define nineteen dissimilarity measures across six families (\S\ref{sec:metrics}).
Measures that we introduce or adapt for proofs are marked \new.
\item We list ten desirable properties (\S\ref{sec:props}).
Eight are tested empirically and two are judged analytically.
\item We build a controlled test bed (\S\ref{sec:case}): five formal proofs of one claim in a single kernel, with a shared lemma library.
Three of the proofs are ``the same'' as one another, one is ``similar'' and one is ``very dissimilar''.
\item We report a similarity table and a property table (\S\ref{sec:results}), together with a sensitivity analysis.
\item We replicate the study in Lean~4/Mathlib on three theorems and nineteen proofs, with all dependency, vocabulary and subject-area information extracted automatically from the library (\S\ref{sec:lean}).
We also discuss what the disagreements between families reveal (\S\ref{sec:discussion}).
\item We relate the problem to six areas of the literature (\S\ref{sec:related}), and set out what a proof metric would enable: checking and evaluating LLM-generated proofs, understanding proofs conceptually, mapping and exploring proof space, and maintaining formal libraries. We also argue that the properties we test are exactly what makes a measure safe to optimise (\S\ref{sec:enable}).
\end{enumerate}
The claim and proofs are classical, chosen so that intuition is uncontroversial.
The value of the exercise is to see which formal measures agree with that intuition, and why some do not.
Non-standard terms, and terms introduced in this paper, are explained in the glossary (Appendix~\ref{app:glossary}).

\input{related_work}

\section{The test bed}\label{sec:case}

\subsection{Claim and proofs}
All proofs establish the same closed first-order sentence
\[
\textsc{Claim}:\qquad \neg\,\exists S.\ \mathrm{Fin}(S)\wedge\big((\forall p.\ p\in S\to \mathrm{Prime}(p))\wedge(\forall p.\ \mathrm{Prime}(p)\to p\in S)\big),
\]
``no finite set is exactly the set of primes''.
We call the three conjuncts \emph{Fin}, \emph{AllPrime} and \emph{Cover}.
The informal texts used by the text-based measure ship with the codepack\footnote{The codepack (code, Lean sources, extracted data, results and the survey instrument) is available at \url{https://github.com/nnikolaou/Proof-space-similarity}.} (\texttt{texts/}).
\begin{description}[leftmargin=1.2em,itemsep=2pt]
\item[E (Euclid, by contradiction).] Assume $S$ is a finite cover of the primes. Put $N=\prod S+1\ge 2$ and let $q=\mathrm{lpf}(N)$, the least prime factor of $N$. By Cover, $q\in S$, so $q\mid\prod S$. Also $q\mid N$, so $q$ divides two consecutive integers, which is absurd.
\item[\Ep{} (Euclid, lemma + corollary).] First prove the lemma ``every finite $T$ misses some prime'', using the same construction. Then derive the claim as a corollary.
\textbf{Intuition: the same proof as E.}
\item[\Epl{} (Euclid, finer-grained).] Identical to E, except that the step $q\in S\Rightarrow q\mid\prod S$ is expanded via $\prod S=q\cdot\prod(S\setminus\{q\})$ and $q\mid q\cdot m$.
\textbf{Intuition: the same proof as E}, written at a finer granularity. We use \Epl{} as a robustness probe.
\item[G (Goldbach, 1730).] The Fermat numbers $F_i=2^{2^i}+1$ are pairwise coprime. If $|S|=k$, then the $k+1$ primes $\mathrm{lpf}(F_0),\dots,\mathrm{lpf}(F_k)$ all lie in $S$. By the pigeonhole principle two of them coincide, contradicting coprimality.
\textbf{Intuition: similar to E.} Both proofs build numbers that are forced to have new prime factors, but they use different constructions and a different combinatorial core.
\item[F (Furstenberg, 1955~\cite{furstenberg}).] In the topology on $\mathbb{Z}$ generated by arithmetic progressions, each $p\mathbb{Z}$ is closed. Hence $\bigcup_{p\in S}p\mathbb{Z}=\mathbb{Z}\setminus\{\pm1\}$ is closed, so $\{\pm1\}$ is open. But a nonempty open set is infinite.
\textbf{Intuition: very dissimilar to E.} The vocabulary is topological rather than arithmetical.
\end{description}

\subsection{The kernel}
The codepack contains a small Curry--Howard kernel (\texttt{proofdist/logic.py}) for intuitionistic first-order natural deduction.
Its connectives are $\to,\wedge,\forall,\exists,\bot$.
Its proof terms are
\[
\begin{aligned}
M ::={}& h \mid c \mid \lambda h{:}A.M \mid M\,N \mid \Lambda x.M \mid M\,t \mid \langle M,N\rangle \mid \pi_1M \mid \pi_2M\\
\mid{}& \mathsf{pack}(t,M) \mid \mathsf{unpack}\,M\,\mathsf{as}\,(x,h)\,\mathsf{in}\,N \mid \mathsf{let}\,h{=}M\,\mathsf{in}\,N,
\end{aligned}
\]
where $c$ ranges over lemma constants of a shared library (Appendix~\ref{app:library}).
Each lemma has a statement, its dependencies, MSC2020 codes and a realizer.
Every proof term is type-checked against \textsc{Claim}.
Reduction is $\beta$, $\forall\beta$, $\wedge\beta$, $\exists\beta$ and $\zeta$ (\textsf{let}-unfolding).
The system is simply typed, so normal forms exist and are unique up to $\alpha$-equivalence.

For example, the normal form of E is (with $N:=\prod S+1$ and $q:=\mathrm{lpf}(N)$)
\[
\begin{aligned}
&\lambda H.\ \mathsf{unpack}\ H\ \mathsf{as}\ (S,h)\ \mathsf{in}\\
&\quad\textsf{prime\_dvd\_succ}\ q\ (\textstyle\prod S)\ P\\
&\qquad\big(\textsf{dvd\_prod}\ S\ q\ (\pi_1h)\ (\pi_2\pi_2 h\ q\ P)\big)\\
&\qquad\big(\textsf{lpf\_dvd}\ N\ (\textsf{prod\_ge2}\ S\ (\pi_1 h))\big),
\end{aligned}
\]
where $P:=\textsf{lpf\_prime}\ N\ (\textsf{prod\_ge2}\ S\ (\pi_1h))$.
The normal form of \Ep{} is \emph{literally the same term} up to $\alpha$-renaming.
Seven reduction steps ($\zeta$, $\forall\beta$, two $\beta$, $\exists\beta$, two $\wedge\beta$) inline the lemma and discharge the detour.
Table~\ref{tab:facts} summarises the five proofs.

\begin{table}[h]
\centering\small
\resizebox{\linewidth}{!}{\input{toy_results/proof_facts.tex}}
\caption{The five formal proofs. $|M|$ and $|\nf(M)|$ are proof-tree sizes (nodes, annotations erased). ``Steps'' is the number of normal-order reduction steps to normal form. ``Lemmas'' counts distinct library constants in the normal form. ``Deps'' is the size of their transitive dependency closure, including foundation nodes. ``Reducts'' is the number of $\alpha$-distinct terms reachable by reduction. ``Nominal strength'' is the strongest foundation node in the dependency closure (\S\ref{sec:metrics}).}
\label{tab:facts}
\end{table}

\subsection{Flow graphs and realizers}
Two pieces of extra machinery support the structural and semantic measures.

\paragraph{Flow tags.} Every atom occurrence (including $\bot$) in the goal, in each lemma instance and in each annotation receives a tag.
Whenever type checking identifies two formulas, for example the argument type with the domain of an implication, the tags of corresponding atoms are merged with union--find.
The resulting classes are the \emph{flow graph} of the derivation: which atom occurrences of the conclusion are ``wired'' to which atom occurrences of which lemmas.
This is a first-order, lemma-relative version of Lambek's and Do\v{s}en's \emph{generality} of a proof~\cite{dosen,lambek}.

\paragraph{Realizers and Herbrand witnesses.} Each lemma comes with a Python realizer, and proof terms are evaluated call-by-value (\texttt{semantics.py}).
A proof of $\neg\exists S\,(\dots)$ is run against a concrete finite set $S$ of primes.
The \emph{Cover} hypothesis is realised by an oracle that answers a query $p$ if $p\in S$ and otherwise raises \textsc{Counterexample}$(p)$.
This is the Kreisel no-counterexample/Herbrand reading of a refutation~\cite{kreisel,kohlenbach}.
Running the proof against a purported finite cover returns a concrete prime that the cover misses, together with the trace of all queries made.
We use all $2^{10}$ subsets of the first ten primes as test sets.

\section{Distances between proofs}\label{sec:metrics}

Every measure maps a pair of proofs to $d\in[0,1]$ (or $\infty$); we report similarity $s=1-d$, with $s=0$ for $d=\infty$ (this arises only for \textsf{Shortest 2-cell}, when two proofs have no common reduct).
We write $J(X,Y)=1-|X\cap Y|/|X\cup Y|$ for the Jaccard distance.
Each measure is named in the form ``\textsf{Name} (\emph{Group})''.

\subsection{Group 1 --- Syntactic}
\begin{description}[leftmargin=1.2em,itemsep=2pt]
\item[\metric{Raw TED}{Syntactic}.]
The Zhang--Shasha tree edit distance~\cite{zhang} between Curry-style proof trees, with unit costs and names kept, normalised by $|T_A|+|T_B|$.
\item[\metric{$\alpha$-TED}{Syntactic}.] The same after $\alpha$-canonical renaming of all binders.
\item[\metric{NF-TED}{Syntactic}\new.] $\alpha$-TED between normal forms. This is the quantitative analogue of the ``same normal form'' criterion of identity.
\item[\metric{Rule-profile}{Syntactic}\new.] The weighted Jaccard distance $1-\sum_r\min(a_r,b_r)/\sum_r\max(a_r,b_r)$ between the multisets of inference rules (term constructors) in the normal forms.
\end{description}

\subsection{Group 2 --- Structural}
\begin{description}[leftmargin=1.2em,itemsep=2pt]
\item[\metric{Flow-linking}{Structural}\new.]
Let $\mathcal L(M)$ be the set of triples (goal atom position, lemma, lemma atom position) that share a flow class in $\nf(M)$. Then $d=J(\mathcal L(A),\mathcal L(B))$.
Two proofs have distance 0 exactly when they wire the conclusion to the same places in the same lemmas.
\item[\metric{Hypothesis-usage}{Structural}\new.] The Jaccard distance between the sets of conjuncts (\emph{Fin}, \emph{AllPrime}, \emph{Cover}) that $\nf(M)$ actually projects from its hypothesis.
A proof that does not use a hypothesis proves a stronger theorem, so this is a cheap proxy for \emph{generality}.
\item[\metric{Signature}{Structural}\new.] The Jaccard distance between the non-logical vocabularies: function symbols in the normal form, plus predicate and function symbols of the lemma statements it uses.
\end{description}

\subsection{Group 3 --- Dependency \& resource}
\begin{description}[leftmargin=1.2em,itemsep=2pt]
\item[\metric{Dependency-Jaccard}{Dependency \& Resource}.] The Jaccard distance between the transitive dependency closures (lemmas plus foundation nodes).
\item[\metric{Dependency-GED}{Dependency \& Resource}.] The label-preserving graph edit distance between the dependency DAGs, $(|V_A\triangle V_B|+|E_A\triangle E_B|)/(|V_A|+|V_B|+|E_A|+|E_B|)$.
\item[\metric{Axiomatic-strength}{Dependency \& Resource}\new.] \mbox{}\\ $d=|\sigma(A)-\sigma(B)|/6$.
Here $\sigma$ is the level of the strongest foundation node used, on the scale
I$\Delta_0$ (0), I$\Delta_0+\Omega_1$ (1), I$\Delta_0+\mathrm{exp}$ (2), RCA$_0$ (3), ACA$_0$ (4), Z$_2$ (5), third-order (6).
This measures \emph{nominal} strength, as written, not the minimal strength needed; see \S\ref{sec:discussion}.
\item[\metric{MSC-EMD}{Dependency \& Resource}\new.] Each proof induces a distribution over MSC2020 codes, with each used lemma's mass split over its codes.
$d$ is the earth mover's distance on the MSC tree (edge weights $1,\tfrac12,\tfrac14$ for the 2-digit, letter and full levels), normalised by the diameter $3.5$.
This is a ``purity of methods'' measure~\cite{purity}.
\item[\metric{Size-ratio}{Dependency \& Resource}.] $1-\min(|\nf A|,|\nf B|)/\max(|\nf A|,|\nf B|)$.
\end{description}

\subsection{Group 4 --- Semantic}
Let $w_M(S)$ be the Herbrand witness and $\tau_M(S)$ the query trace of $M$ on test set $S$, and let $\mathcal S$ be the family of test sets.
\begin{description}[leftmargin=1.2em,itemsep=2pt]
\item[\metric{Herbrand-witness}{Semantic}\new.] $\frac1{|\mathcal S|}\sum_S[w_A(S)\ne w_B(S)]$: the fraction of inputs on which the two proofs exhibit different missing primes.
\item[\metric{Query-trace}{Semantic}\new.] The mean normalised Levenshtein distance between $\tau_A(S)$ and $\tau_B(S)$. This compares the whole interaction with the cover, not just its outcome.
\item[\metric{Witness-magnitude}{Semantic}\new.] The mean of $|\ln w_A(S)-\ln w_B(S)|/\max(\ln w_A(S),\ln w_B(S))$. This is a proof-mining-style comparison of the bounds the proofs yield.
\end{description}

\subsection{Group 5 --- Information-theoretic}
\begin{description}[leftmargin=1.2em,itemsep=2pt]
\item[\metric{NCD-term}{Information-theoretic}.] The normalised compression distance~\cite{cilibrasi} $\mathrm{NCD}(x,y)=\frac{C(xy)-\min(C(x),C(y))}{\max(C(x),C(y))}$, with zlib at level 9.\footnote{zlib is the standard, freely available compression library also used in ZIP, gzip and PNG. Its DEFLATE algorithm replaces repeated substrings by back-references and then encodes the remaining symbols with variable-length (Huffman) codes; level 9 is its strongest compression setting. Compressed size $C(\cdot)$ serves as a practical stand-in for the uncomputable Kolmogorov complexity.}
It is applied to the $\alpha$-canonical serialisations of the normal forms.
NCD is a computable stand-in for the normalised information distance $\max(K(x|y),K(y|x))/\max(K(x),K(y))$~\cite{li}.
\item[\metric{NCD-text}{Information-theoretic}.] NCD between the informal proof texts.
\end{description}
We also report the directed quantity $\hat K(B|A)=(C(AB)-C(A))/C(B)$, which estimates how much knowing $A$ helps in writing down $B$.

\subsection{Group 6 --- Higher-categorical}
\begin{description}[leftmargin=1.2em,itemsep=2pt]
\item[\metric{Shortest 2-cell}{Higher-categorical}\new.] Treat proofs as 1-cells and reduction steps as 2-cells. Let $L$ be the length of the shortest ``valley'' zig-zag $A\to^{*}R\,{}^{*}\!\!\leftarrow B$, found by exhaustive BFS over the reducts of each proof ($L=\infty$ if none exists). Then $d=L/(1+L)$.
\item[\metric{Identification-level}{Higher-categorical}\new.] \mbox{}\\ A ladder of identification criteria, from finest to coarsest:
$R_0$ $\alpha$-equivalence;
$R_1$ convertibility;
$R_2$ same flow-linking (generality);
$R_3$ same Herbrand content;
$R_4$ same conclusion (propositional truncation).
Let $\equiv_k$ be the equivalence relation generated by $R_0\cup\dots\cup R_k$ on the universe of proofs considered. Then $d(A,B)=\min\{k: A\equiv_k B\}/4$.
The $\equiv_k$ are nested by construction, so $d$ is an \emph{ultrametric}.
$R_2$ and $R_3$ are not comparable in general, so placing generality below computational content is a modelling choice. It reads ``at which level of truncation do $A$ and $B$ become the same proof?''~\cite{hott}.
\end{description}

\section{Desirable properties}\label{sec:props}
Let $\mathcal U=\{\mathrm E,\mathrm{\Ep},\mathrm{\Epl},\mathrm G,\mathrm F\}$.
Properties P1--P8 are tested by the code; P9 and P10 are analytic judgements, justified in \texttt{properties.py}.
\begin{description}[leftmargin=2.6em,itemsep=1pt]
\item[P1] \emph{Reflexivity}: $d(X,X)=0$ on $\mathcal U$.
\item[P2] \emph{Symmetry}: $d(X,Y)=d(Y,X)$ on $\mathcal U$.
\item[P3] \emph{Triangle inequality} on all triples of $\mathcal U$.
\item[P4] \emph{$\alpha$-invariance}: $d(\mathrm E,\alpha\mathrm E)=0$ for an $\alpha$-renamed copy.
\item[P5] \emph{Bureaucracy invariance}: $d(\mathrm E,\mathrm{\Ep})\le 0.05$, i.e.\ insensitivity to inessential presentational differences such as extracting a lemma or introducing local abbreviations.
\item[P6] \emph{Granularity robustness}: $d(\mathrm E,\mathrm{\Epl})<\min(d(\mathrm E,\mathrm G),d(\mathrm E,\mathrm F))$, i.e.\ a finer encoding of E stays closer to E than any different proof.
\item[P7] \emph{Discrimination}: $d(\mathrm E,\mathrm G)>0.05$ and $d(\mathrm E,\mathrm F)>0.05$ (\pmark{} if only one holds).\footnote{The threshold $0.05$, here and in P5, is an implementation choice. The intended conditions are $d>0$ for P7 and $d=0$ for P5; the margin keeps measures with small numerical noise (e.g.\ compression-based ones, for which even identical inputs give $d>0$) from passing or failing on noise alone. With a strict $d>0$, no P7 mark in either study would change. For P5 the margin matters: in the Lean study, \textsf{Module-EMD} and, for some theorems, \textsf{Vocabulary}, \textsf{Rule-profile} and \textsf{Size-ratio} pass only because their distances for the ``same'' pair are small but non-zero ($\le 0.05$).}
\item[P8] \emph{Intuitive ordering}: $d(\mathrm E,\mathrm{\Ep})<d(\mathrm E,\mathrm G)<d(\mathrm E,\mathrm F)$ (\pmark{} if ``same'' is closest but G and F are mis-ordered or tied).
\item[P9] \emph{Effective computability}: decidable and feasible on realistic proofs.
\item[P10] \emph{Informal applicability}: can be evaluated on proofs as mathematicians write them.
\end{description}

\section{Results}\label{sec:results}

\begin{table}[p]
\centering\small
\resizebox{\linewidth}{!}{\input{toy_results/similarity_main.tex}}
\caption{\textbf{Similarity scores} $s=1-d$ for the three intuitive pairs. ``Order'' is \cmark{} if $s(\mathrm E,\mathrm{\Ep})>s(\mathrm E,\mathrm G)>s(\mathrm E,\mathrm F)$, \pmark{} if only the ``same'' pair comes out on top, and \xmark{} otherwise. The last column shows the robustness probe (E, \Epl), which is also intuitively ``the same''. Measures marked \new{} in \S\ref{sec:metrics} are new or adapted. Generated by \texttt{run\_all.py}; all ten pairwise distances are in Appendix~\ref{app:allpairs}.}
\label{tab:sim}
\end{table}

\begin{figure}[p]
\centering
\includegraphics[width=0.92\linewidth]{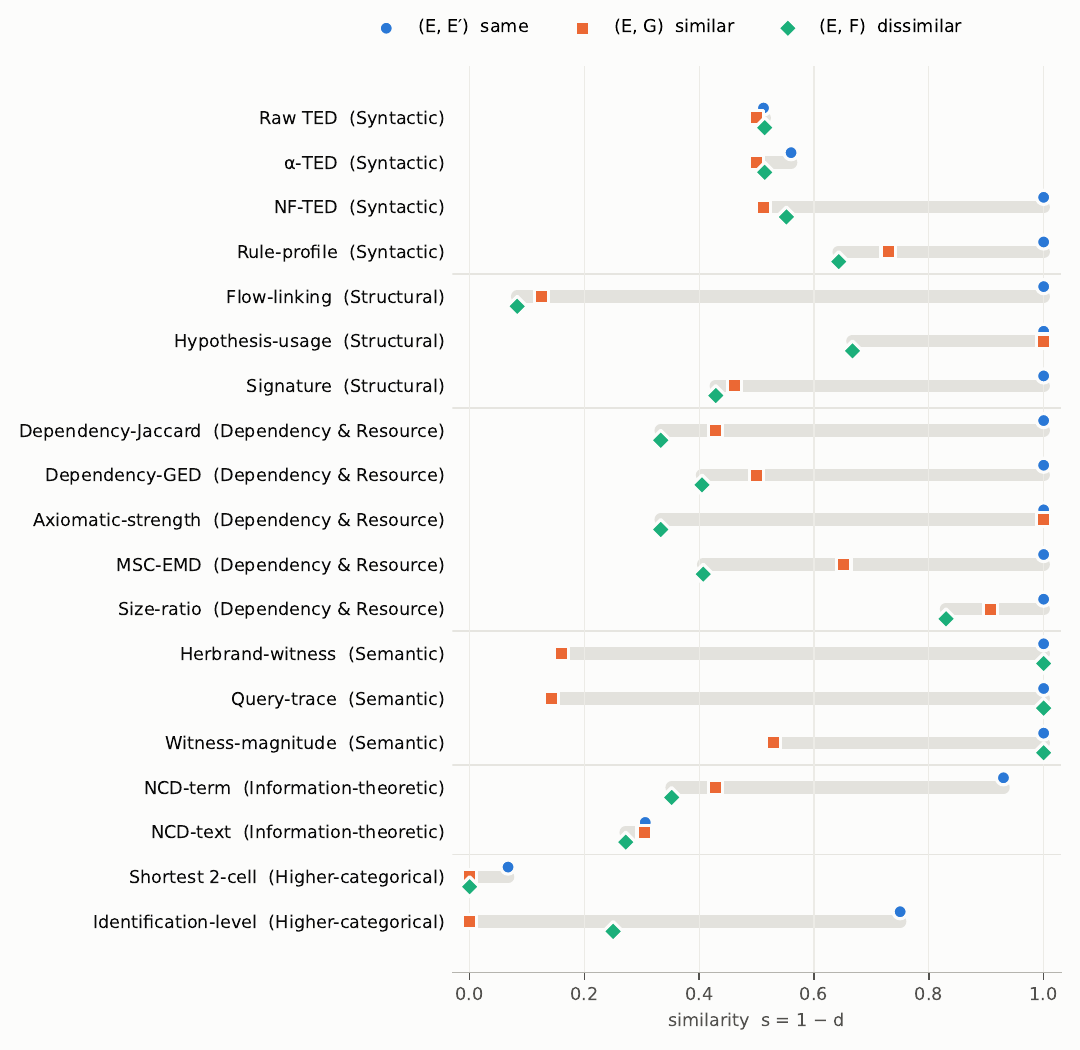}
\caption{The data of Table~\ref{tab:sim} as a dot plot. Each row is one measure. A measure agrees with intuition when the blue circle (same) lies right of the orange square (similar), which lies right of the green diamond (dissimilar). The semantic rows show the inversion: Furstenberg's proof sits with Euclid's.}
\label{fig:sim}
\end{figure}

\begin{table}[t]
\centering\small
\resizebox{\linewidth}{!}{\input{toy_results/properties.tex}}
\caption{\textbf{Desirable properties} (\S\ref{sec:props}). \cmark{} satisfied, \xmark{} violated, \pmark{} partially or with caveats. ``Score'' counts \cmark{} as 1 and \pmark{} as $\tfrac12$. P1--P8 are computed by the code (evidence in \texttt{results/property\_evidence.json}); P9--P10 are analytic.}
\label{tab:props}
\end{table}

\subsection{Similarity}
Table~\ref{tab:sim} and Figure~\ref{fig:sim} give the scores.
\begin{itemize}[leftmargin=*,itemsep=2pt]
\item \textbf{Raw syntax cannot see sameness.} Raw TED scores (E,\Ep) at $0.512$, essentially the same as (E,G) at $0.500$ and (E,F) at $0.514$. Lemma extraction and \textsf{let}-bindings cost as many edits as a different argument. $\alpha$-canonicalisation barely helps.
\item \textbf{Normalisation identifies E and \Ep{} exactly} (NF-TED $=1$), but then fails to separate G from F: $0.512$ against $0.552$, so Furstenberg comes out \emph{closer} to Euclid than Goldbach does.
\item \textbf{Structural and dependency measures reproduce intuition.} Flow-linking, Signature, Dependency-Jaccard, Dependency-GED, MSC-EMD, Rule-profile, Size-ratio and NCD-term all give the full ordering. Flow-linking separates most sharply ($1.000$ / $0.125$ / $0.083$).
\item \textbf{Coarse resource measures cannot separate E from G.} Hypothesis-usage and Axiomatic-strength give $s(\mathrm E,\mathrm G)=1$. Only Furstenberg uses \emph{AllPrime} (to know $1\notin p\mathbb Z$ and that $p\mathbb Z$ is closed), and only Furstenberg, as written, quantifies over families of sets.
\item \textbf{Every semantic measure inverts intuition.} The Herbrand witness of F equals that of E on all $1024$ test sets, $w(S)=\mathrm{lpf}(1+\prod S)$; for instance $w(\{2,3,5,7\})=211$ for both. Goldbach's proof queries $\mathrm{lpf}(F_0),\mathrm{lpf}(F_1),\dots$ in order and returns the first one missing ($17$ for $\{2,3,5,7\}$, with trace $3,5,17$). So semantically E$\,\equiv\,$F$\,\not\equiv\,$G.
\item \textbf{NCD-text is dominated by wording.} The ordering is right, but the margins are tiny ($0.306$ / $0.304$ / $0.272$), and the prose rewrite \Ep{} scores \emph{lower} than the near-verbatim \Epl{} ($0.830$).
\item \textbf{Higher-categorical measures are discrete.} Shortest 2-cell is finite only for (E,\Ep), with $L=14$. Identification-level places (E,\Ep) at convertibility (0.75), (E,F) at ``same Herbrand content'' (0.25) and (E,G) at the top (0). This is again the semantic inversion, now built into an ultrametric.
\end{itemize}

\subsection{Properties}
Table~\ref{tab:props} gives the property marks.
No measure satisfies all ten.
The best overall scores (9.5/10) go to \textsf{Dependency-Jaccard}, \textsf{Dependency-GED} and \textsf{MSC-EMD}, followed by \textsf{Signature} (9). They lose points only on P10 (and, for Signature, P9): informal use requires extracting the cited results or vocabulary by hand.
\textsf{Flow-linking}, \textsf{Rule-profile} and \textsf{Signature} also order correctly and are invariant, but they depend on normal forms, which hurts P9 and P10.
Robustness to granularity (P6) is the property most often violated by measures that otherwise behave well. Size-ratio, Shortest 2-cell, Identification-level, every semantic measure and both coarse resource measures fail it, and the structural measures pass with less margin.
The two NCD measures fail reflexivity, symmetry and $\alpha$-invariance, as is well known for compression-based estimators on short strings.

\begin{table}[h]
\centering\small
\input{toy_results/directed_compression.tex}
\caption{Directed conditional compression $\hat K(B|A)=(C(AB)-C(A))/C(B)$. Lower values mean that knowing $A$ makes $B$ cheaper to write. In both representations $\hat K(\mathrm E|\mathrm F)<\hat K(\mathrm F|\mathrm E)$: the longer proof F supplies much of what is needed to write E, but not conversely. For (E,\Epl), knowing \Epl{} makes E almost free ($0.061$), reflecting that \Epl{} \emph{refines} E.}
\label{tab:dir}
\end{table}

\subsection{Sensitivity of the semantic identification}
The witness agreement between E and F depends on how the lemma ``a nonempty open set contains arbitrarily large elements'' is realised.
Our realizer returns the \emph{first} element $x+b$ of the progression $x+b\mathbb Z$ that the openness witness provides. At $x=1$, with modulus $b=\mathrm{lcm}(S)=\prod S$, this is $1+\prod S$.
If it is replaced by the \emph{second} element $x+2b$, Furstenberg's witness becomes $\mathrm{lpf}(1+2\prod S)$, for example $13$ instead of $7$ for $S=\{2,3\}$, and \metric{Herbrand-witness}{Semantic} between E and F jumps from $0.000$ to $1.000$ (\texttt{results/sensitivity.json}).
The first element is the canonical choice, and the agreement matches the known observation that Furstenberg's argument ``is Euclid's in disguise''~\cite{mercer}.
Still, semantic identity of proofs that use non-constructive lemmas is only as determinate as the realizers chosen for those lemmas.

\section{Replication in Lean~4 with Mathlib}\label{sec:lean}

The kernel study served as a useful example to introduce our framework, but it has two weaknesses: the proofs were hand-formalised against a hand-built lemma library, and there is only one theorem.
We therefore repeat the exercise in Lean~4~\cite{lean4} against a pinned revision of Mathlib~\cite{mathlib} (\texttt{lakefile.toml}), where the library, the elaborator and the kernel are not of our making.

\subsection{Theorems and proofs}
Table~\ref{tab:leanproofs} lists the nineteen proofs (plus one $\alpha$-renamed copy of each base proof, used only for property P4).
The roles ``same'', ``finer'', ``similar'' and ``dissimilar'' are the author's stipulation; \S\ref{sec:lean-survey} suggests validating them with human ratings as the next step.
Proofs are written as a working Lean user would write them, i.e.\ mostly in tactic mode, so the elaborated term is produced by Lean and not by us.

\begin{table}[h]
\centering\small
\begin{tabular}{lllp{8.4cm}}
\toprule
Theorem & Proof & Role & Idea \\
\midrule
Primes & E & base & Euclid: least prime factor of $\prod S+1$, by contradiction\\
\emph{no finite $S$ is} & E$'$ & same & Euclid as a lemma (``every finite set of primes misses one'') plus corollary\\
\emph{the set of primes} & E$^+$ & finer & E with $q\mid\prod S$ via $\prod S=q\prod(S\setminus q)$ and $q\mid1$ via $q\le1<2\le q$\\
 & M & similar & Mathlib's \texttt{Nat.exists\_infinite\_primes} (least prime factor of $n!+1$)\\
 & G & similar & Goldbach: Fermat numbers are pairwise coprime, plus pigeonhole\\
 & F & dissimilar & Furstenberg: the arithmetic-progression topology on $\mathbb Z$, built from scratch\\
 & Eu & dissimilar & Euler/Erd\H{o}s: $\sum_p 1/p$ diverges (Mathlib's \texttt{Nat.Primes.not\_summable\_one\_div})\\
\addlinespace
$\sqrt2$ irrational & P & base & parity and strong induction (infinite descent)\\
\emph{$a^2=2b^2\Rightarrow b=0$} & P$'$ & same & P as two lemmas: $2\mid n^2\Rightarrow2\mid n$, and the descent step\\
 & P$^+$ & finer & P with $2\mid n^2\Rightarrow 2\mid n$ by an even/odd case split and explicit cancellations\\
 & T & similar & Tennenbaum's descent $(a,b)\mapsto(2b-a,\,a-b)$\\
 & V & similar & parity of the 2-adic valuation (\texttt{Nat.factorization})\\
 & R & dissimilar & Mathlib's \texttt{irrational\_sqrt\_two} over $\mathbb R$\\
\addlinespace
Gauss's sum & I & base & induction, closed by \texttt{ring}\\
\emph{$2\sum_{i\le n}i=n(n+1)$} & I$'$ & same & explicit recursor \texttt{Nat.rec} with the step as a lemma\\
 & I$^+$ & finer & induction with every rewrite justified individually (no \texttt{ring})\\
 & Gp & similar & Gauss pairing: add the sum to its reflection\\
 & H & dissimilar & hockey-stick identity $\sum_i\binom i1=\binom{n+1}2$\\
 & B & dissimilar & Faulhaber's formula with Bernoulli numbers\\
\bottomrule
\end{tabular}
\caption{Lean/Mathlib proofs (\texttt{ProofDist/*.lean}). Each theorem also has an $\alpha$-variant of its base proof (E$\alpha$, P$\alpha$, I$\alpha$).}
\label{tab:leanproofs}
\end{table}

\subsection{What is extracted}
A Lean metaprogram (\texttt{ProofDist/Exporter.lean}) records, for every proof constant:
\begin{itemize}[leftmargin=*,itemsep=1pt]
\item the \emph{elaborated proof term} as a Curry-style tree: implicit and instance arguments and binder types are erased, bound variables are de~Bruijn indices (so the tree is $\alpha$-invariant by construction);
\item a \emph{local normal form}: proof-local constants (auxiliary lemmas, \texttt{match} auxiliaries, the Furstenberg topology, tactic-generated lemmas) are $\delta$-unfolded, and $\beta$-, $\zeta$- (\texttt{let}/\texttt{have}) and projection-of-constructor redexes are contracted, with a step count.
Mathlib constants are \emph{not} unfolded; normalising through the library would be the ``invariance under lemma refinement'' discussed in \S\ref{sec:discussion}, and is infeasible at Mathlib scale;
\item the constants occurring in the normal form, each with its kind (theorem, definition, inductive, \dots), its Mathlib module, and whether it is a type class;
\item the axioms the proof depends on, and the head symbols of the types of all $\lambda$-binders;
\item the \emph{transitive dependency closure} of the proof in the Mathlib environment (types and values of every constant reached), together with the dependency edges;
\item the result of asking the kernel whether every pair of proofs of the same theorem is definitionally equal.
\end{itemize}
The source text of each proof (comments removed) is read from the \texttt{.lean} files.

\subsection{Measures}
The eighteen Lean measures keep the six families and, where possible, the definitions of \S\ref{sec:metrics}; differences are deliberate.
\begin{description}[leftmargin=1.2em,itemsep=2pt]
\item[Syntactic.] \textsf{Source-edit}: token-level Levenshtein distance between the source texts, normalised by the longer one. \textsf{Term-TED}, \textsf{NF-TED}: Zhang--Shasha distance between elaborated and locally-normalised Curry trees (pq-gram distance~\cite{pqgram} when both trees are very large; flagged $^\dagger$). \textsf{Rule-profile}: weighted Jaccard distance between multisets of node kinds of the normal form.
\item[Structural.] \textsf{Vocabulary}: Jaccard distance between the non-theorem constants (definitions, inductives, constructors) in the normal form. \textsf{Typeclass-footprint}\new: the same for type classes, i.e.\ the algebraic/topological structures the proof invokes. \textsf{Binder-profile}\new: weighted Jaccard distance between multisets of binder-type heads (what the proof quantifies over).
\item[Dependency \& resource.] \textsf{Direct-premises}: Jaccard distance between the Mathlib theorems used. \textsf{Dependency-Jaccard}, \textsf{Dependency-GED}: as before, on the full Mathlib closure (the GED uses that closures are downward-closed, so $|E_A\triangle E_B|=\sum_{u\in V_A\triangle V_B}\mathrm{outdeg}(u)$). \textsf{Axiom-profile}\new: Jaccard distance between axiom sets (\texttt{propext}, \texttt{Classical.choice}, \texttt{Quot.sound}), a coarse constructive/classical signal replacing nominal strength. \textsf{Module-EMD}\new: earth mover's distance on Mathlib's module hierarchy (edge weight $2^{-\ell}$ at depth $\ell$) between the distributions of the proof's direct constants over modules. This replaces hand-assigned MSC codes with the library's own taxonomy. \textsf{Size-ratio}: as before.
\item[Semantic.] Witness extraction is not available: Lean proofs of propositions are computationally irrelevant, and most use \texttt{Classical.choice}. We report \textsf{Kernel-defeq}, the identity the Lean kernel itself uses; by definitional proof irrelevance it is $0$ for every pair, which is precisely the ``propositional truncation'' level of \S\ref{sec:metrics}. The Herbrand measures of the kernel study remain our semantic evidence.
\item[Information-theoretic.] \textsf{NCD-source} and \textsf{NCD-term} (on the serialised normal form).
\item[Higher-categorical.] \textsf{Shortest 2-cell}: total reduction steps if two proofs have the same local normal form ($0$ if their terms already coincide), $\infty$ otherwise. \textsf{Identification-level}: ladder $R_0$ identical term, $R_1$ same local normal form, $R_2$ same direct Mathlib premises, $R_3$ same Mathlib areas (module prefixes of depth 3), $R_4$ kernel definitional equality; $d=\min\{k: A\equiv_k B\}/4$.
\end{description}
Properties P1--P8 are evaluated per theorem (with the theorem's proofs as $\mathcal U$; for P6--P8 ``similar'' and ``dissimilar'' range over all proofs with that role) and marked \cmark{} if they hold for all three theorems, \xmark{} if for none, \pmark{} otherwise.

\subsection{Results}
Table~\ref{tab:leanfacts} summarises the proofs; Table~\ref{tab:leansim} gives similarity to the base proof, averaged over the three theorems; Table~\ref{tab:leanprops} gives the property marks.
Per-theorem tables are in Appendix~\ref{app:lean}.

\begin{table}[h]
\centering\small
\resizebox{\linewidth}{!}{\input{lean_results/lean_proof_facts.tex}}
\caption{Lean proofs: source tokens, sizes of the elaborated and locally-normalised Curry trees ($^*$ = normaliser fuel exhausted), number of Mathlib theorems used directly, size of the dependency closure, and axioms.}
\label{tab:leanfacts}
\end{table}

\begin{table}[h]
\centering\small
\resizebox{\linewidth}{!}{\input{lean_results/lean_similarity_pooled.tex}}
\caption{\textbf{Lean/Mathlib: similarity to the base proof}, averaged over the three theorems (for ``similar'' and ``dissimilar'', first averaged over the proofs with that role). ``Order'' is \cmark{} if same $>$ similar $>$ dissimilar holds for every theorem, \pmark{} if for some theorem or if ``same'' is always closest; $k/3$ counts theorems in full order. $^\dagger$: pq-gram approximation used for some pairs.}
\label{tab:leansim}
\end{table}

\begin{table}[h]
\centering\small
\resizebox{\linewidth}{!}{\input{lean_results/lean_properties.tex}}
\caption{\textbf{Lean/Mathlib: desirable properties} (\S\ref{sec:props}), aggregated over the three theorems.}
\label{tab:leanprops}
\end{table}

\paragraph{Overview.}
All nineteen proofs and the three $\alpha$-variants compiled against the pinned Mathlib, and the kernel confirmed that all 70 within-theorem pairs (including $\alpha$-variants) are definitionally equal.
Dependency closures range from about 3{,}400 to 6{,}500 Mathlib constants for the elementary proofs. They jump to 11{,}930 (R), 15{,}681 (B) and 16{,}247 (Eu) for the proofs that cite analytic results.
Normal forms range from 7 nodes (Eu, a two-line citation) to 12{,}836 (P, dominated by the certificates of \texttt{nlinarith}).
For pairs where both trees are large (mostly the $\sqrt2$ proofs), TED was approximated by pq-gram distance.
For F, the normaliser's fuel ran out before the end, so its normal form is only partly reduced.

\paragraph{Replication.}
Five measures reproduce the full intuitive ordering (same $>$ similar $>$ dissimilar) on \emph{all three} theorems: \textsf{NF-TED} ($0.81/0.40/0.14$), \textsf{Rule-profile}, \textsf{Dependency-Jaccard} ($1.00/0.66/0.45$), \textsf{Dependency-GED} ($1.00/0.75/0.52$) and \textsf{NCD-term}.
Two measures score best on the properties (9/10): \textsf{Dependency-Jaccard} and \textsf{Dependency-GED}, which were also among the best in the kernel study.
They give the ``same'' proofs (E$'$, P$'$, I$'$) \emph{identical} dependency closures, they separate all other proofs, and they satisfy P1--P5 and P7--P8.
They lose half a point each on P6: in the primes theorem the finer Euclid E$^+$ is placed further from E than Goldbach's proof is, because \texttt{Finset.mul\_prod\_erase} pulls in a different part of the library.
\textsf{Module-EMD} (8.5), \textsf{Vocabulary} and \textsf{Direct-premises} (8 each) follow.
The finding of \S\ref{sec:results} that dependency-based measures track intuition best therefore survives moving from a toy library to Mathlib, from one theorem to three, and from hand-written proof terms to elaborated tactic proofs.

\paragraph{Pre-registered expectations.}
\begin{enumerate}[label=(\roman*),leftmargin=*,itemsep=2pt]
\item \emph{Source- and term-level measures separate less well than NF-TED}: confirmed.
\textsf{Source-edit} ranks ``dissimilar'' above ``similar'' on average ($0.17$ vs $0.19$) and is in full order for only one theorem. \textsf{Term-TED} is in full order for one theorem.
The elaborated term is dominated by automation: in P the \texttt{nlinarith} certificates are inlined, whereas in P$'$ they sit inside auxiliary lemmas. Hence \textsf{Term-TED}$(\mathrm P,\mathrm P')$ is $0.04$, while after normalisation it is $0.82$.
\item \emph{Citation proofs stress dependency measures}: confirmed, with a twist.
Eu, R and B are the furthest proofs in their theorems under \textsf{Dependency-Jaccard} ($0.33$, $0.33$, $0.24$), as expected from their huge closures.
But M, which cites Mathlib's own Euclid-style proof ($n!+1$), is placed \emph{further} from E than Goldbach's proof is ($0.66$ vs $0.88$), and so is it by \textsf{NF-TED} ($0.46$ vs $0.62$).
A proof that delegates its content to a library lemma looks different from a proof that spells out the same idea, at every level we measure.
Proof by citation thus behaves like an extreme form of the granularity problem: the argument is the same, but it sits in someone else's lemma.
\item \emph{Module-EMD places F, Eu, R and B furthest}: only partly confirmed.
It does so for the primes (F $0.61$, Eu $0.53$), but not for $\sqrt2$, where V (via \texttt{Nat.factorization}) is further than R, nor for Gauss's sum, where B ($0.85$) is \emph{closer} than the pairing and hockey-stick proofs.
Inspection shows why. The direct constants of every proof are dominated by bookkeeping lemmas from \texttt{Init}, \texttt{Mathlib.Algebra} and \texttt{Mathlib.Tactic} (casts, \texttt{simp} and \texttt{ring} lemmas). Furstenberg's topology, built from scratch inside the proof, contributes no \texttt{Mathlib.Topology} constants at all, and B's analytic content is hidden behind one cited lemma.
A post-hoc check (Table~\ref{tab:posthoc}, not pre-registered) shows that IDF weighting does not fix this, but computing the area profile over the whole dependency closure restores the average ordering on all three theorems, though with a compressed range.
\item \emph{Kernel-defeq fails P6--P8}: confirmed. Lean's built-in identity is propositional truncation.
\end{enumerate}

\paragraph{Findings specific to Lean.}
\begin{itemize}[leftmargin=*,itemsep=2pt]
\item \emph{No ``same'' pair is convertible.} In the kernel study, E and E$'$ had the same normal form. In Lean, none of E/E$'$, P/P$'$ or I/I$'$ do, so \textsf{Shortest 2-cell} is uninformative and \textsf{NF-TED} fails P5.
Tactics such as \texttt{rintro}/\texttt{obtain} and \texttt{induction} versus an explicit \texttt{Nat.rec} elaborate to structurally different terms. Equality of normal forms, the classical criterion for identity of proofs, is too fine for proofs as actually written in a proof assistant.
The identity that does survive is at the level of dependencies: E/E$'$ and P/P$'$ meet at ``same direct premises'' in the identification ladder.
\item \emph{Axioms carry no information.} Every proof, including the induction proof of Gauss's sum, depends on \texttt{propext}, \texttt{Classical.choice} and \texttt{Quot.sound}, because Mathlib's tactics and basic API use them throughout.
\textsf{Axiom-profile} is therefore constant. Constructivity, which matters for the semantic level and for questions such as whether a disproof of ETH yields an algorithm (\S\ref{sec:events}), cannot be read off \texttt{\#print axioms} in Mathlib practice. It needs a finer analysis of \emph{where} choice is used.
\item \emph{Auto-generated auxiliary lemmas.} Lean abstracts sub-proofs into auxiliary constants named after the parent theorem (e.g.\ \texttt{PD.Primes.Ealpha.\_proof\_1\_1}). Before we canonicalised these names, raw-term measures were not $\alpha$-invariant. This is a small but instructive example of a property (P4) failing for reasons that have nothing to do with mathematics.
\end{itemize}

\begin{table}[h]
\centering\small
\input{lean_results/lean_posthoc_module_idf.tex}
\caption{Post-hoc sensitivity analysis of the area measure (not pre-registered): similarity to the base proof under \textsf{Module-EMD} on direct constants (as in Table~\ref{tab:leansim}), with IDF weighting and bookkeeping modules removed, and computed over the whole dependency closure (Mathlib modules truncated to depth 3).}
\label{tab:posthoc}
\end{table}

\paragraph{Rank agreement with the stipulated roles (post hoc).}
As a further check, not pre-registered, Table~\ref{tab:spearman} gives the Spearman correlation between each measure's similarity-to-base and the stipulated roles. For this, the roles are encoded on an ordinal scale: 3 for proofs that are intuitively the same as the base proof (both ``same'' and ``finer'', since a finer-grained version is still the same argument), 2 for ``similar'' and 1 for ``dissimilar'' proofs.
Within theorems, \textsf{NCD-term} ranks proofs best (mean $\rho=0.91$), followed by \textsf{NF-TED} ($0.83$) and \textsf{Dependency-Jaccard} ($0.82$).
Pooled across theorems, \textsf{Vocabulary} is best ($\rho=0.81$; $0.85$ without the finer proofs), followed by \textsf{NF-TED} and \textsf{Direct-premises} ($0.76$).
The two views answer different questions. The per-theorem mean asks whether a measure ranks a theorem's proofs correctly. The pooled value also asks whether its scale means the same across theorems. \textsf{NCD-term} ranks well within each theorem but its scale shifts between them, which lowers its pooled value to $0.52$; the dependency measures are affected similarly but less.
Several caveats apply.
\begin{itemize}[leftmargin=*,itemsep=1pt]
\item The samples are tiny (5--6 proofs per theorem) and heavily tied, so per-theorem values near $0.96$ are close to the maximum attainable and differences between them carry little weight.
\item The permutation $p$-values are one-sided, exploratory and uncorrected for testing fifteen measures.
\item The reference ordering is the author's stipulation, so the table measures agreement with that stipulation, not with expert judgement. The rating study of \S\ref{sec:lean-survey} would turn it into a genuine validation.
\end{itemize}
Within these limits, the table corroborates the main finding (normal-form and dependency measures track the intuitive ordering best), and it shows that \textsf{Vocabulary} performs better on this criterion than its property score suggests.

\begin{table}[h]
\centering\small
\resizebox{\linewidth}{!}{\input{lean_results/lean_spearman_roles.tex}}
\caption{Post hoc, not pre-registered: Spearman correlation between similarity-to-base and the stipulated roles (same $=$ finer $=3$, similar $=2$, dissimilar $=1$), per theorem ($n=6,5,5$), as the mean of the three, pooled over theorems ($n=16$, with a one-sided permutation $p$-value, uncorrected), and pooled without the finer proofs ($n=13$). Bold marks the highest value in each column; ``--'' means undefined because the measure is constant. Generated by \texttt{analysis/spearman\_roles.py}.}
\label{tab:spearman}
\end{table}

\subsection{Suggested next step: human ratings}\label{sec:lean-survey}
The roles used throughout (same, finer, similar, dissimilar) are the author's judgement, so every ``intuitive ordering'' result above is conditional on them.
We suggest validating them with a rating study as the next step.
The codepack already contains a ready-to-use instrument (\texttt{survey/}): all 51 within-theorem pairs, in random order and orientation, with one-paragraph informal sketches and a 1--7 scale (1 = the same proof, 7 = completely different proofs).
With ratings from a panel of mathematicians, one would report inter-rater agreement and the Spearman correlation of each measure with the median rating. This would replace stipulated roles with empirical ground truth, and would show whether experts' notion of ``same proof'' is itself a single scale or, as our results suggest, several.

\section{Discussion}\label{sec:discussion}

\paragraph{Different families measure different things.}
The families do not disagree at random. Each one examines proofs under a different notion of ``inessential difference'':
\begin{itemize}[leftmargin=*,itemsep=1pt]
\item syntactic measures ignore nothing, or only bureaucracy once normalised;
\item structural measures ignore everything except wiring and vocabulary;
\item dependency measures ignore everything except \emph{what is cited};
\item semantic measures ignore everything except \emph{what is computed}.
\end{itemize}
From the semantic viewpoint, Furstenberg's topology is scaffolding around Euclid's number, which is what Mercer showed by hand~\cite{mercer}.
Intuition, by contrast, seems to weigh \emph{concepts used}, which is why the dependency and MSC measures track it best.
In our test bed this gives a sharp answer to ``how dissimilar can two proofs be?''.
Two proofs can be maximally far apart conceptually (MSC-EMD similarity $0.41$, flow-linking similarity $0.08$) and yet \emph{identical in computational content}.
Conversely, two proofs in closely related vocabulary (E and G) can compute different things.

\paragraph{Granularity is the hard problem.}
After normalisation, the main threat is not bureaucracy but \emph{granularity}: the same argument written against a finer lemma library looks different.
\Epl{} shows this. It changes one lemma into three, and NF-TED similarity drops from $1$ to $0.78$ and Flow-linking similarity to $0.43$.
A satisfactory metric probably needs either a canonical library, which is what large formal libraries such as Mathlib approximate, or invariance under \emph{lemma refinement}. That means treating a lemma and a proof of it as interchangeable, i.e.\ normalising through the library as well as through the proof.

\paragraph{Nominal and minimal strength.}
Axiomatic-strength separates F from E only \emph{as written}.
Furstenberg's sets are all $\Delta_0$-definable and the topology can be eliminated. Indeed, the infinitude of primes is provable in I$\Delta_0+\mathrm{exp}$, and whether I$\Delta_0$ alone suffices is a famous open problem~\cite{pww}.
The \emph{minimal} strength of each proof (the weakest system in which that proof, suitably translated, still goes through) would put all five proofs at the same level.
In general it is not computable.

\paragraph{Which measures to use.}
Taking both studies together (Tables~\ref{tab:sim}, \ref{tab:props}, \ref{tab:leansim} and \ref{tab:leanprops}), the measures fall into four groups.
\begin{itemize}[leftmargin=*,itemsep=2pt]
\item \emph{Strongest single scores: \textsf{Dependency-Jaccard} and \textsf{Dependency-GED}.}
They score best on the properties in both studies (9.5/10 in the kernel, 9/10 in Lean). They reproduce the intuitive ordering on every theorem, treat the ``same'' proofs as identical, and are cheap to compute without normalisation.
Their known weakness is proof by citation: a proof that delegates its argument to a library lemma looks far from the same argument spelled out (M versus E).
\item \emph{Good rankers, poor identity tests: \textsf{NF-TED} and \textsf{NCD-term}.}
They give the full ordering on all Lean theorems with the widest same-to-dissimilar spread (\textsf{NF-TED} $0.81 \to 0.14$). However, they never identify differently written versions of one argument, because tactic elaboration differs, and NCD also fails reflexivity and symmetry.
\item \emph{Promising but fragile: area measures (\textsf{MSC-EMD}, \textsf{Module-EMD}).}
They score best in the kernel study, where subject codes were assigned by hand. When computed from Mathlib modules, routine lemmas dominate them unless the profile is taken over the whole dependency closure (Table~\ref{tab:posthoc}).
\item \emph{Weak as distances.} Some measures are degenerate in Lean: \textsf{Kernel-defeq} (zero for every pair, by proof irrelevance), \textsf{Axiom-profile} (constant across Mathlib) and \textsf{Shortest 2-cell} (no two Lean proofs share a normal form). Others mostly track wording and automation rather than arguments: \textsf{Source-edit}, raw/term TED and text-level NCD. \textsf{Typeclass-footprint} and \textsf{Hypothesis-usage} are binary and register only large changes of method.
\end{itemize}
The semantic Herbrand measures need a separate verdict. By agreement with intuition they are the worst, since they invert the ordering. By information they are among the best, since they alone show that Furstenberg's proof computes Euclid's witness. They do not measure \emph{less} of the same thing as the other families; they measure something else.

\begin{figure}[p]
\centering
\includegraphics[height=0.9\textheight,width=\linewidth,keepaspectratio]{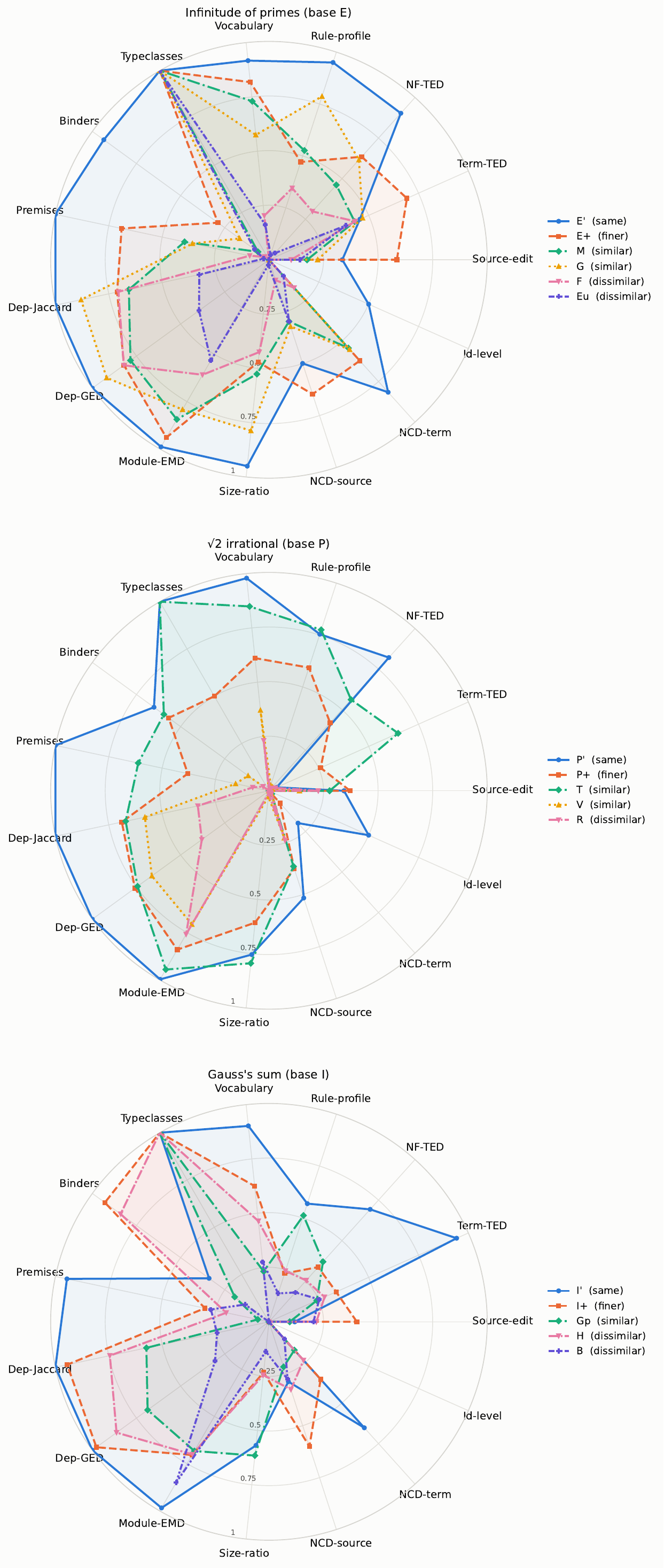}
\caption{Proof profiles in the Lean/Mathlib study (top: infinitude of primes, base E; middle: $\sqrt2$, base P; bottom: Gauss's sum, base I). Each axis is one measure, and each polygon gives one proof's similarity to its theorem's base proof (outer = more similar). Colours and line styles encode the stipulated role (same, finer, similar, dissimilar) consistently across panels. Measures that are constant in this study (Kernel-defeq, Axiom-profile, Shortest 2-cell) are omitted. The shapes depend on the order of the axes, so the figure illustrates the profiles qualitatively; exact values are in Appendix~\ref{app:lean}. Generated by \texttt{analysis/make\_radar.py}.}
\label{fig:radar}
\end{figure}

\paragraph{Towards a profile, not a number.}
This classification is the strongest argument for the thesis of this paper: no single measure does everything well. The ones that fail as distances are not useless; they mark \emph{levels} at which proofs coincide or part company.
We therefore propose to report the dissimilarity of two proofs as a profile with three parts:
\begin{enumerate}[leftmargin=*,itemsep=1pt]
\item the \emph{identification level} at which they meet (same term, same normal form, same premises, same computational content, same area, or only the same theorem);
\item a \emph{within-level distance} that ranks pairs at that level: \textsf{NF-TED} among proofs with comparable normal forms, \textsf{Dependency-GED} across them;
\item a small set of \emph{orthogonal coordinates} that capture what the ladder hides: an area distance (closure-based \textsf{Module-EMD}), a semantic distance where witnesses can be extracted, the size of the dependency closure, which flags proof by citation, and an \emph{independence flag}, which records whether a proof rests on a library version of the theorem it proves (see below).
\end{enumerate}
On our data such profiles read as a mathematician would describe the pairs:
\begin{itemize}[leftmargin=*,itemsep=1pt]
\item (E,\Ep) in the kernel are convertible. In Lean they share all their premises but not their normal forms.
\item (E,F) agree computationally but are far apart in concepts and dependencies.
\item (E,G) differ computationally, yet are close in dependencies and area.
\item (E,M) make the same argument, but at different levels of citation.
\end{itemize}
No single scalar here achieves this.
Figure~\ref{fig:radar} shows the same point visually. Across the measures, the ``same'' proofs trace the outer polygon, and citation proofs such as Eu and B collapse to small, irregular shapes. No single axis separates the roles, but the overall \emph{shape} of each proof's profile does.

Applications that need a number, such as ranking, deduplication or a novelty reward for agents, can still derive one, for example a lexicographic order on (level, within-level distance) or a weighted sum. They should do so knowingly, choosing the components that satisfy the properties their use requires (Table~\ref{tab:uses}). A profile makes it much harder to game a measure than a single score does, because a proof must differ at several levels at once to count as new.

\paragraph{Circularity and independence.}
Can a proof in our studies be circular, and do the measures notice?
Lean rules out \emph{literal} circularity. A declaration may only use earlier declarations, so the environment is acyclic and no theorem can depend on itself. Recursive proofs are accepted only as genuine induction. The only way to assume a conclusion outright is an \texttt{axiom} or \texttt{sorry}, which the exporter's axiom record would expose; none of our proofs contains either.
(In the kernel study, library lemmas are axioms, so a lemma restating the claim would \emph{not} be caught automatically; ours were checked by inspection.)
What Lean does not rule out is \emph{relative} circularity: proving a theorem from a library result whose own proof depends on that theorem.
To detect it, we listed by hand the Mathlib statements that trivially imply each of our claims, and searched each proof's dependency closure for them (\texttt{analysis/independence\_check.py}; Table~\ref{tab:indep}).
Two hits are expected: M cites \texttt{Nat.exists\_infinite\_primes} and R cites \texttt{irrational\_sqrt\_two}, so both prove their theorem from the library's own version of it.
The third is not.
Mathlib's proof that $\sum_p 1/p$ diverges, on which our Euler/Erd\H{o}s proof Eu rests, itself depends on the infinitude of primes:
\begin{center}\small
\texttt{Nat.Primes.not\_summable\_one\_div} $\to$ \texttt{not\_summable\_one\_div\_on\_primes} $\to$ \texttt{one\_half\_le\_sum\_primes\_ge\_one\_div} $\to$\\
\texttt{Nat.smoothNumbersUpTo\_card\_le} $\to\cdots\to$ \texttt{Nat.mem\_smoothNumbers'} $\to$ \texttt{Nat.mem\_factoredNumbers'} $\to$ \texttt{Nat.exists\_infinite\_primes}.
\end{center}
Nothing is logically wrong: the library's proof of \texttt{Nat.exists\_infinite\_primes} is Euclid's, which is not circular. But as an \emph{alternative} proof, Eu is not independent of Euclid in Mathlib. It is Euclid plus a great deal more, and the dependency enters through a technical lemma about smooth numbers, probably for convenience rather than necessity.
None of our measures factors this out. The dependency measures contain the edge but weigh it like any of several thousand others, and term-level measures do not look inside library lemmas at all.
Relative circularity is thus an extreme form of proof by citation, and it motivates adding \emph{independence} to the profile.
Call two proofs of a theorem independent if neither's closure contains a library equivalent of the theorem or the other's key lemmas. By this test Eu is not independent of E, while G and F are.
Such a check matters for AI-generated proofs, which can rely silently on the result they claim to prove, or on an equivalent statement, in ways that neither the kernel nor \texttt{\#print axioms} reveals. It also matters for claims of a ``new, independent'' proof.
For informal proofs, detecting that a proof begs the question requires first extracting the results it cites, which is a statement-level problem.

\begin{table}[h]
\centering\small
\input{lean_results/lean_independence.tex}
\caption{Relative circularity: whether a proof's Mathlib dependency closure contains a library statement that trivially implies the theorem being proved, and the length of the shortest dependency chain to it (0 = cited directly). The list of library equivalents was compiled by hand.}
\label{tab:indep}
\end{table}

\paragraph{Beyond a single theorem.}
So far, all our comparisons have been between proofs of the \emph{same} claim, but most measures never look at the statement. They compare proofs as objects (terms, dependency sets, vocabularies, compressed strings), so they apply unchanged to proofs of \emph{different} claims. What changes is the meaning of the result. Within a theorem the statement is fixed, so every difference is a difference of method. Across theorems, \emph{statement} similarity (the claims are about related objects) and \emph{method} similarity (the arguments work the same way) are entangled.
The interesting cross-theorem case is ``the same trick on a different theorem'', for instance infinite descent for $\sqrt2$ and for $x^4+y^4=z^2$, and a raw score confuses it with ``the same objects, a different trick''.
The measures transfer to different degrees:
\begin{itemize}[leftmargin=*,itemsep=1pt]
\item \textsf{Rule-profile}, \textsf{Binder-profile}, \textsf{Typeclass-footprint}, \textsf{Module-EMD} and \textsf{Axiom-profile} transfer directly, as signatures of how a proof is built rather than what it proves.
\item \textsf{NF-TED}, \textsf{NCD-term} and \textsf{Flow-linking} transfer once statement-specific constants are abstracted away (replaced by their kind, or compared up to renaming).
\item Dependency measures transfer but are dominated by shared infrastructure unless ubiquitous lemmas are discounted (e.g.\ by IDF weighting).
\item Semantic measures need a common interface, for example both proofs producing a prime.
\item \textsf{Kernel-defeq} and the top rung of the identification ladder (``same theorem'') lose their meaning, and the ladder needs new levels such as ``same proof template'' and ``related statements''.
\end{itemize}
An exploratory comparison of all 120 cross-theorem pairs of our Lean proofs (Table~\ref{tab:cross}; post hoc) bears this out.
Every measure rates proofs of the same theorem as more similar on average than proofs of different theorems. For \textsf{Dependency-Jaccard} the gap is small ($0.58$ versus $0.50$): shared Mathlib infrastructure swamps the signal.
The most similar cross-theorem pairs reveal kinship of \emph{style} rather than of idea. By dependencies, Eu ($\sum_p 1/p$ diverges) and B (Faulhaber) are among the closest ($0.75$), being both citation proofs that invoke analytic machinery. By \textsf{Rule-profile}, the explicit-rewriting proofs E$^+$ and Gp are closest ($0.85$).
None of the current measures reliably detects ``the same idea across theorems''.

A natural dedicated measure is \emph{anti-unification}~\cite{plotkin,reynolds}: compute the most specific common generalisation of two (abstracted) proof terms, the largest template of which both proofs are instances, and use its size relative to the two proofs as a similarity.
It is zero for unrelated proofs and close to one for the same template applied to different statements. It also comes with an explanation, the template itself, which can often be extracted as a new library lemma.
Cross-theorem similarity would support mapping families of methods across a library (descent, pigeonhole, Euclid-style constructions, induction schemes) and analogy and transfer between results. It would also support detecting ``a new theorem, an old trick'' in AI-generated mathematics, where whether a proof instantiates a known method matters as much as whether its result is new, and comparing related but distinct results such as the forced and unforced Euler and Navier--Stokes theorems (\S\ref{sec:events}).

\begin{table}[h]
\centering\small
\resizebox{\linewidth}{!}{\input{lean_results/lean_cross_theorem.tex}}
\caption{Exploratory, post hoc: mean similarity between proofs of the same theorem (within, 51 pairs) and of different theorems (across, 120 pairs) for the Lean proofs ($\alpha$-variants excluded), and the most similar cross-theorem pair under each measure. Generated by \texttt{analysis/cross\_theorem.py}.}
\label{tab:cross}
\end{table}

\paragraph{Threats to validity.}
(i) The proofs are hand-formalised at a chosen granularity, and the lemmas are axioms with declared dependencies, not formalised down to foundations.
(ii) MSC codes and foundation levels are author judgements.
(iii) The sample is five proofs of one theorem, so the empirical triangle and symmetry checks are necessary conditions only.
(iv) Compression on strings under 1\,kB is noisy.
(v) Herbrand content is sampled on $1024$ inputs, not proved equal.
Section~\ref{sec:lean} addresses (i) and (ii) by moving to Lean~4/Mathlib, at the price of losing the semantic measures; (iii) is reduced to three theorems and nineteen proofs, which is still small.

\input{applications}

\section{Future work}\label{sec:future}
Below are some ideas for immediate future work based on the principles described in this paper.

\paragraph{Human ground truth (next step).}
The roles same/similar/dissimilar are our stipulation. The rating study prepared in \S\ref{sec:lean-survey} (51 pairs, 1--7 scale) would test them, measure inter-rater agreement, and correlate every measure with expert judgement.
A natural extension asks raters to score the appraisal dimensions studied by Inglis and Aberdein~\cite{inglis}: simplicity, explanatoriness and purity. Hilbert's original question about \emph{simplicity} could then be put to an empirical test.

\paragraph{Scale.}
\begin{itemize}[leftmargin=*,itemsep=1pt]
\item \emph{Theorems with many formal proofs}. Alternative proofs can be collected from Mathlib's \texttt{Archive}, from its version history, and from the outputs of neural provers on benchmarks such as miniF2F~\cite{minif2f}. The latter offer many proofs of each statement, though dominated by automation.
\item \emph{All of Mathlib}. Cross-theorem similarity and duplicate detection at this scale need a compiled, streaming exporter, compact sketches of dependency closures (MinHash) with locality-sensitive hashing to generate candidate pairs, a fast tree-edit-distance implementation, and manual validation of a sample.
\item \emph{Machine-generated corpora}, such as \texttt{openai/math}~\cite{oai-math}. About 42\% of its top-line results are formalised in Lean, and the exporter applies to them. For the informal remainder, citations and subject areas would have to be extracted with language-model assistance and validated, and each family placed relative to the human literature.
\end{itemize}

\paragraph{Comparing independently formalised results.}
The 2026 Euler and Navier--Stokes formalisations (\S\ref{sec:events}) are natural test cases. OpenAI's two are public in one repository, and they can be compared directly (cross-theorem, at the levels of dependencies, vocabulary and lemma structure).
Comparing either with the independent forced-Euler formalisation of Buckmaster and Alp\"oge would need access to their Lean sources (currently not available).

\paragraph{Granularity invariance.}
The main open technical problem is to compare a proof that cites a lemma with one that spells it out. That requires normalising \emph{through} the library, treating a lemma and its proof as interchangeable and quotienting by lemma refinement, at a cost that remains feasible at library scale.

\paragraph{Automatic independence checks.}
The independence check of \S\ref{sec:discussion} relies on a hand-made list of library equivalents of each theorem. Automating it would make it usable at library scale and on AI-generated proofs. Library constants whose statements imply the claim could be found by trying to close the claim from each candidate with a short, time-limited tactic call, or by statement-similarity retrieval. A proof would then be reported as circular relative to the library, or as independent, together with the shortest chain. The same machinery would also check whether a lemma found by an agent is merely a restatement of the goal.

\paragraph{Cross-theorem similarity.}
Extend the measures to proofs of different claims, separating statement similarity from method similarity (\S\ref{sec:discussion}). The principled route is anti-unification of abstracted proof terms, with the size of the common template as the similarity and the template as the explanation. Applied to Mathlib (combined with the scaling work above), this would yield a map of proof methods across the library, and a test of whether AI-generated results are new theorems proved by old tricks.

\paragraph{Semantics in real libraries.}
Lean's proof irrelevance and Mathlib's pervasive classical axioms hide computational content. Locating where choice is \emph{essentially} used, and extracting witnesses from constructive fragments by proof mining~\cite{kohlenbach}, would bring the Herbrand level, under which Furstenberg's proof is Euclid's, to real libraries. It would also decide questions such as whether a formal disproof of ETH yields an algorithm.

\paragraph{AI-driven theorem proving.}
\begin{itemize}[leftmargin=*,itemsep=1pt]
\item \emph{Orchestration of prover agents.} In multi-agent systems, a proof-dissimilarity score can steer the swarm. Agents are assigned so that their partial \emph{proof trajectories} stay mutually dissimilar, which maximises their complementarity, and effort spent rediscovering the same route is reduced. Quality-diversity search over proof space and diversity-aware rewards are concrete instances to evaluate.
\item \emph{Evaluation.} Report distinct-proof@$k$ alongside pass@$k$, measure contamination by nearest-neighbour distance to training corpora, and track how training changes a prover's position in proof space.
\item \emph{Interpretability.} For an AI-generated proof, retrieve the \emph{most similar human-understandable proofs}, of the same or related theorems, and align them with the generated proof. Where the two agree, the familiar proof explains the unfamiliar one. Where they diverge, the divergence localises what is genuinely new and needs human attention.
\end{itemize}

\paragraph{Theory.}
Characterise which combinations of properties P1--P8 are jointly satisfiable, give invariance theorems for the identification ultrametric, and study the geometry of proof space: its diameter on bureaucracy-free proofs, and its dependence on the library.

\section{Conclusion}
Mathematics is entering a period in which proofs are produced faster than people can read them. AI systems now generate research-level arguments in large numbers, and a proof assistant's kernel can certify that each is correct. Correctness, however, is not understanding. Deciding whether a proof is genuinely new, how many distinct ideas a batch of generated proofs contains, how much a machine added to what humans supplied, and how an unfamiliar argument relates to familiar ones are all questions about how far apart proofs are. Hilbert's long-neglected 24th problem has thus become a practical one, and this paper has tried to make it measurable.

The question ``how far apart can two proofs of the same theorem be?'' has a different answer for each notion of ``essential'' content.
On a controlled test bed, conceptual and dependency measures reproduce mathematicians' intuition.
Computational content tells a different and, as Mercer observed, correct story: Furstenberg's proof is Euclid's.
Raw syntax tells no story at all.
We suggest that future work treat proof dissimilarity as a \emph{multi-level profile}, with granularity invariance as the central open requirement (\S\ref{sec:future}).
The Lean/Mathlib replication confirms on three theorems that dependency-closure measures track intuition best, and shows that equality of normal forms is too fine an identity criterion for proofs written with tactics. It also shows that most of the profile can be computed automatically from a large formal library, and that a proof assistant's own notion of identity (definitional proof irrelevance) sits at the coarsest level of the profile, so it cannot by itself say how different two proofs are.

\section*{Code and data availability}
All code, the Lean~4/Mathlib sources and exporter, the extracted proof data, the scripts that regenerate every table and figure, and the human-rating survey instrument are available at \url{https://github.com/nnikolaou/Proof-space-similarity}. Code is released under the MIT licence and data under CC~BY~4.0.

\section*{Statement on the use of AI}
This work was carried out with substantial assistance from the large language model Claude (Anthropic). Under the author's direction, it proposed candidate measures and properties, produced the code and assisted in analyzing the results and drafting the manuscript. The author posed the research question, directed and reviewed the work and edited the draft. He takes full responsibility for the content. All formal proofs are machine-checked (by the codepack's type checker and by the Lean~4 kernel against a pinned Mathlib), every reported number is regenerated from the released code, and analyses added after the Lean results were obtained are marked as post hoc.


\appendix
\section{Lemma library}\label{app:library}
\begin{center}\small
\input{toy_results/library.tex}
\end{center}
Foundation nodes and their strength levels: Logic, PA\_basic, IDelta0 (0); Exp, FiniteSetCoding (2); SecondOrderSets (3); FamiliesOfSets (6).
Internal dependency nodes (e.g.\ \texttt{lpf\_def}, \texttt{topology\_ax}) are listed in \texttt{library.py}.

\section{All pairwise distances}\label{app:allpairs}
\begin{center}\scriptsize
\resizebox{\linewidth}{!}{\input{toy_results/distance_all_pairs.tex}}
\end{center}

\section{Lean/Mathlib: per-theorem similarities}\label{app:lean}
Similarity $s=1-d$ of each proof to its theorem's base proof.

\begin{center}\small
\textbf{Primes} (base E)\\[2pt]
\resizebox{0.9\linewidth}{!}{\input{lean_results/lean_similarity_primes.tex}}\\[10pt]
\textbf{$\sqrt2$} (base P)\\[2pt]
\resizebox{0.9\linewidth}{!}{\input{lean_results/lean_similarity_sqrt2.tex}}\\[10pt]
\textbf{Gauss's sum} (base I)\\[2pt]
\resizebox{0.9\linewidth}{!}{\input{lean_results/lean_similarity_gauss.tex}}
\end{center}

\input{glossary}

\section{Reproducing the results}
Kernel study (standard-library Python):
\begin{verbatim}
cd toy && python3 run_all.py && python3 make_figure.py
\end{verbatim}
Lean/Mathlib study (Lean toolchain via elan; Mathlib revision pinned in lakefile.toml):
\begin{verbatim}
scripts/build.sh               # cache get + lake build
scripts/export_and_analyse.sh  # export, metrics, paper
\end{verbatim}
\end{document}

%% file: related_work.tex
\section{Related work}\label{sec:related}
Proof similarity sits at the meeting point of several literatures that rarely cite each other.
We group them by the question each one asks about proofs.

\paragraph{When are two proofs the same?}
Hilbert's unpublished 24th problem asked for a criterion of proof simplicity, and in effect of proof identity~\cite{thiele}. Thiele and Wos connected it to automated reasoning and to the search for shorter, simpler proofs with resolution provers~\cite{thielewos}.
General proof theory took up the identity question after Prawitz's \emph{normalisation conjecture}: two natural-deduction derivations represent the same proof iff they have the same normal form~\cite{prawitz,kreisel}.
The categorical tradition reads proofs as morphisms in free categories with structure, and identity as equality of morphisms~\cite{lambek,lambekscott}. Coherence theorems then tell us when a formula has essentially one proof~\cite{dosenpetric}.
Do\v{s}en combined normalisation with \emph{generality}, the pattern of which atom occurrences are linked, as a second criterion~\cite{dosen}. Wideb\"ack's thesis gives a systematic treatment of identity in this setting~\cite{wideback}.
For classical logic, where cut elimination is not confluent and normalisation identifies too much, the main candidates are proof nets~\cite{girard}, Hughes's combinatorial proofs~\cite{hughes} and the atomic flows of deep inference~\cite{strassburger,gugliemiflows}.
Homotopy type theory turns identity of proofs of an equation into a structured space of paths~\cite{hott}; our Identification-level ultrametric borrows its truncation levels.
All of these give equivalence relations. Our contribution is to grade them: \textsf{NF-TED}, \textsf{Flow-linking} and \textsf{Identification-level} are quantitative versions of normalisation, generality and truncation respectively.

\paragraph{How costly is a proof?}
Proof complexity measures proofs by size within a fixed proof system~\cite{cookreckhow,krajicek}. Gödel's speed-up theorem shows that the shortest proofs of one statement in two systems can differ by any computable function~\cite{goedel36}.
Reverse mathematics measures the axioms a theorem needs~\cite{simpson}. Our \textsf{Axiomatic-strength} and \textsf{Axiom-profile} apply the same idea to individual proofs rather than theorems.
The infinitude of primes is a well-known test case: whether it is provable in I$\Delta_0$ is still open~\cite{pww}.
Proof mining reads proofs for their computational content~\cite{kohlenbach}. Our Herbrand measures compare two proofs by the content they yield, which is how Furstenberg's proof is revealed as Euclid's~\cite{mercer}.

\paragraph{Why prove a theorem again?}
The philosophy of mathematical practice studies the many proofs of one theorem as a phenomenon in itself.
Topics include why mathematicians re-prove results~\cite{dawson}, what proofs are for beyond certification~\cite{rav}, purity of methods~\cite{purity}, explanatory proofs~\cite{steiner,lange}, and the methodological content of proofs~\cite{avigad}.
Collections such as Lemmermeyer's catalogue of reciprocity proofs~\cite{lemmermeyer} and \emph{Proofs from THE BOOK}~\cite{aigner} classify proofs by method, which is an informal similarity judgement.
Inglis and Aberdein measured mathematicians' proof appraisals empirically and found that they cluster on a few dimensions~\cite{inglis}. This is the methodology our survey instrument (\S\ref{sec:lean-survey}) aims to extend from appraising one proof to comparing two.

\paragraph{Similarity inside proof assistants and in machine learning.}
Large formal libraries made proofs into data.
Premise selection ranks library facts by similarity to the current goal, using hand-crafted features, kernel methods or neural encoders~\cite{alama,kaliszykurban,mash,deepmath}.
ML4PG clusters Coq proofs by statistical features to find reusable proof patterns~\cite{ml4pg}. Concept alignment across libraries matches definitions by the properties proved about them~\cite{gauthier}.
Proof repair adapts an existing proof to a changed library, and so needs to know which proofs are similar enough to reuse~\cite{qed}.
These works use similarity as an \emph{instrument} for retrieval, clustering or repair. They do not ask which notions of similarity are principled, or how they relate to the identity-of-proofs literature, which is our question.

\paragraph{Neural theorem proving.}
Language-model provers now generate formal proofs at scale: GPT-f~\cite{gptf}, HyperTree Proof Search~\cite{htps}, Draft--Sketch--Prove~\cite{dsp}, LeanDojo~\cite{leandojo}, DeepSeek-Prover~\cite{deepseekprover}, AlphaGeometry~\cite{alphageometry} and AlphaProof~\cite{alphaproof}. Benchmarks such as miniF2F~\cite{minif2f} and autoformalisation~\cite{autoformalization} support them.
In informal mathematics, LLM-generated proofs that are fluent but wrong are a documented problem~\cite{proofbluff}.
Evaluation in this area is almost entirely by pass rate. Whether two generated proofs are different, or whether a proof was merely recalled, is rarely measured. Section~\ref{sec:enable} argues that proof metrics fill this gap.

\paragraph{Program and code similarity.}
Under Curry--Howard, proof similarity is a special case of program similarity, which has a large literature on clone detection~\cite{roy} and learned similarity of code~\cite{codebleu}.
Two differences matter.
First, program similarity usually cares about behaviour, while proofs of propositions are behaviourally irrelevant (Lean's kernel identifies them all).
Second, mathematicians' notion of ``the same argument'' is coarser than syntactic clone types and finer than extensional equality, which is why we need a graded family of measures.
Tree edit distance~\cite{zhang,pqgram} and compression distance~\cite{li,bennett,cilibrasi} are inherited from this literature.

%% file: toy_results/proof_facts.tex
\begin{tabular}{lccccccl}
\toprule
Proof & $|M|$ & $|\mathrm{nf}(M)|$ & steps & lemmas & deps & reducts & nominal strength\\
\midrule
E & 74 & 78 & 7 & 5 & 13 & 128 & I$\Delta_0$+exp\\
E$'$ & 94 & 78 & 7 & 5 & 13 & 10 & I$\Delta_0$+exp\\
E$^{+}$ & 119 & 121 & 8 & 7 & 15 & 256 & I$\Delta_0$+exp\\
G & 92 & 86 & 3 & 6 & 17 & 8 & I$\Delta_0$+exp\\
F & 101 & 94 & 7 & 10 & 23 & 128 & third-order\\
\bottomrule
\end{tabular}

%% file: toy_results/similarity_main.tex
\begin{tabular}{l ccc c c}
\toprule
 & \multicolumn{3}{c}{similarity $s=1-d$} & & robustness\\
\cmidrule(lr){2-4}\cmidrule(lr){6-6}
Metric (Group) & (E,E$'$) & (E,G) & (E,F) & order & (E,E$^{+}$)\\
 & \emph{same} & \emph{similar} & \emph{dissimilar} & & \emph{same}\\
\midrule
Raw TED (Syntactic) & 0.512 & 0.500 & 0.514 & \xmark & 0.756\\
$\alpha$-TED (Syntactic) & 0.560 & 0.500 & 0.514 & \pmark & 0.756\\
NF-TED (Syntactic) & 1.000 & 0.512 & 0.552 & \pmark & 0.779\\
Rule-profile (Syntactic) & 1.000 & 0.729 & 0.643 & \cmark & 0.833\\
\addlinespace
Flow-linking (Structural) & 1.000 & 0.125 & 0.083 & \cmark & 0.429\\
Hypothesis-usage (Structural) & 1.000 & 1.000 & 0.667 & \xmark & 1.000\\
Signature (Structural) & 1.000 & 0.462 & 0.429 & \cmark & 0.727\\
\addlinespace
Dependency-Jaccard (Dependency \& Resource) & 1.000 & 0.429 & 0.333 & \cmark & 0.750\\
Dependency-GED (Dependency \& Resource) & 1.000 & 0.500 & 0.405 & \cmark & 0.839\\
Axiomatic-strength (Dependency \& Resource) & 1.000 & 1.000 & 0.333 & \xmark & 1.000\\
MSC-EMD (Dependency \& Resource) & 1.000 & 0.652 & 0.407 & \cmark & 0.845\\
Size-ratio (Dependency \& Resource) & 1.000 & 0.907 & 0.830 & \cmark & 0.645\\
\addlinespace
Herbrand-witness (Semantic) & 1.000 & 0.160 & 1.000 & \xmark & 1.000\\
Query-trace (Semantic) & 1.000 & 0.142 & 1.000 & \xmark & 1.000\\
Witness-magnitude (Semantic) & 1.000 & 0.530 & 1.000 & \xmark & 1.000\\
\addlinespace
NCD-term (Information-theoretic) & 0.930 & 0.429 & 0.352 & \cmark & 0.728\\
NCD-text (Information-theoretic) & 0.306 & 0.304 & 0.272 & \cmark & 0.830\\
\addlinespace
Shortest 2-cell (Higher-categorical) & 0.067 & 0.000 & 0.000 & \pmark & 0.000\\
Identification-level (Higher-categorical) & 0.750 & 0.000 & 0.250 & \pmark & 0.250\\
\bottomrule
\end{tabular}

%% file: toy_results/properties.tex
\begin{tabular}{lccccccccccc}
\toprule
Metric (Group) & P1 & P2 & P3 & P4 & P5 & P6 & P7 & P8 & P9 & P10 & score\\
\midrule
Raw TED (Syntactic) & \cmark & \cmark & \cmark & \xmark & \xmark & \cmark & \cmark & \xmark & \cmark & \xmark & 6\\
$\alpha$-TED (Syntactic) & \cmark & \cmark & \cmark & \cmark & \xmark & \cmark & \cmark & \pmark & \cmark & \xmark & 7.5\\
NF-TED (Syntactic) & \cmark & \cmark & \cmark & \cmark & \cmark & \cmark & \cmark & \pmark & \pmark & \xmark & 8\\
Rule-profile (Syntactic) & \cmark & \cmark & \cmark & \cmark & \cmark & \cmark & \cmark & \cmark & \pmark & \xmark & 8.5\\
\addlinespace
Flow-linking (Structural) & \cmark & \cmark & \cmark & \cmark & \cmark & \cmark & \cmark & \cmark & \pmark & \xmark & 8.5\\
Hypothesis-usage (Structural) & \cmark & \cmark & \cmark & \cmark & \cmark & \xmark & \pmark & \xmark & \pmark & \pmark & 6.5\\
Signature (Structural) & \cmark & \cmark & \cmark & \cmark & \cmark & \cmark & \cmark & \cmark & \pmark & \pmark & 9\\
\addlinespace
Dependency-Jaccard (Dependency \& Resource) & \cmark & \cmark & \cmark & \cmark & \cmark & \cmark & \cmark & \cmark & \cmark & \pmark & 9.5\\
Dependency-GED (Dependency \& Resource) & \cmark & \cmark & \cmark & \cmark & \cmark & \cmark & \cmark & \cmark & \cmark & \pmark & 9.5\\
Axiomatic-strength (Dependency \& Resource) & \cmark & \cmark & \cmark & \cmark & \cmark & \xmark & \pmark & \xmark & \pmark & \pmark & 6.5\\
MSC-EMD (Dependency \& Resource) & \cmark & \cmark & \cmark & \cmark & \cmark & \cmark & \cmark & \cmark & \cmark & \pmark & 9.5\\
Size-ratio (Dependency \& Resource) & \cmark & \cmark & \cmark & \cmark & \cmark & \xmark & \cmark & \cmark & \pmark & \pmark & 8\\
\addlinespace
Herbrand-witness (Semantic) & \cmark & \cmark & \cmark & \cmark & \cmark & \xmark & \pmark & \xmark & \pmark & \pmark & 6.5\\
Query-trace (Semantic) & \cmark & \cmark & \cmark & \cmark & \cmark & \xmark & \pmark & \xmark & \pmark & \pmark & 6.5\\
Witness-magnitude (Semantic) & \cmark & \cmark & \cmark & \cmark & \cmark & \xmark & \pmark & \xmark & \pmark & \pmark & 6.5\\
\addlinespace
NCD-term (Information-theoretic) & \xmark & \xmark & \cmark & \xmark & \xmark & \cmark & \cmark & \cmark & \cmark & \xmark & 5\\
NCD-text (Information-theoretic) & \xmark & \xmark & \cmark & \xmark & \xmark & \cmark & \cmark & \cmark & \cmark & \cmark & 6\\
\addlinespace
Shortest 2-cell (Higher-categorical) & \cmark & \cmark & \cmark & \cmark & \xmark & \xmark & \cmark & \pmark & \pmark & \xmark & 6\\
Identification-level (Higher-categorical) & \cmark & \cmark & \cmark & \cmark & \xmark & \xmark & \cmark & \pmark & \pmark & \xmark & 6\\
\bottomrule
\end{tabular}

%% file: toy_results/directed_compression.tex
\begin{tabular}{lcccc}
\toprule
Pair $(A,B)$ & $\hat K(B\mid A)$ text & $\hat K(A\mid B)$ text & $\hat K(B\mid A)$ term & $\hat K(A\mid B)$ term\\
\midrule
(E,E$'$) & 0.694 & 0.683 & 0.070 & 0.070\\
(E,G) & 0.696 & 0.621 & 0.571 & 0.500\\
(E,F) & 0.728 & 0.683 & 0.648 & 0.468\\
(E,E$^{+}$) & 0.170 & 0.061 & 0.272 & 0.101\\
\bottomrule
\end{tabular}

%% file: lean_results/lean_proof_facts.tex
\begin{tabular}{llcccccl}
\toprule
Theorem & Proof (role) & tokens & $|M|$ & $|\mathrm{nf}|$ & premises & closure & axioms\\
\midrule
primes & E (base) & 211 & 62 & 287 & 33 & 5435 & propext, choice, sound\\
primes & E' (same) & 204 & 28 & 302 & 33 & 5435 & propext, choice, sound\\
primes & E+ (finer) & 348 & 107 & 609 & 38 & 3920 & propext, choice, sound\\
primes & M (similar) & 56 & 31 & 151 & 20 & 3667 & propext, choice, sound\\
primes & G (similar) & 184 & 104 & 226 & 39 & 6145 & propext, choice, sound\\
primes & F (dissimilar) & 869 & 116 & 122* & 88 & 5842 & propext, choice, sound\\
primes & Eu (dissimilar) & 50 & 21 & 7 & 8 & 16247 & propext, choice, sound\\
sqrt2 & P (base) & 164 & 6385 & 12836 & 114 & 5030 & propext, choice, sound\\
sqrt2 & P' (same) & 215 & 154 & 9673 & 114 & 5030 & propext, choice, sound\\
sqrt2 & P+ (finer) & 358 & 1222 & 7796 & 93 & 4542 & propext, choice, sound\\
sqrt2 & T (similar) & 237 & 5695 & 16169 & 123 & 3424 & propext, choice, sound\\
sqrt2 & V (similar) & 51 & 66 & 343 & 34 & 6499 & propext, choice, sound\\
sqrt2 & R (dissimilar) & 109 & 158 & 152 & 41 & 11930 & propext, choice, sound\\
gauss & I (base) & 32 & 862 & 862 & 37 & 3803 & propext, choice, sound\\
gauss & I' (same) & 153 & 818 & 1514 & 35 & 3803 & propext, choice, sound\\
gauss & I+ (finer) & 72 & 201 & 201 & 15 & 3592 & propext, choice, sound\\
gauss & Gp (similar) & 189 & 212 & 531 & 44 & 3985 & propext, choice, sound\\
gauss & H (dissimilar) & 106 & 215 & 211 & 17 & 4591 & propext, choice, sound\\
gauss & B (dissimilar) & 93 & 4209 & 6274 & 108 & 15681 & propext, choice, sound\\
\bottomrule
\end{tabular}

%% file: lean_results/lean_similarity_pooled.tex
\begin{tabular}{l ccc cc c}
\toprule
 & \multicolumn{3}{c}{mean similarity to base} & & & robustness\\
\cmidrule(lr){2-4}\cmidrule(lr){7-7}
Metric (Group) & same & similar & dissimilar & order & $k/3$ & finer\\
\midrule
Source-edit (Syntactic) & 0.266 & 0.168 & 0.186 & \pmark & 1 & 0.453\\
Term-TED$^\dagger$ (Syntactic) & 0.479 & 0.341 & 0.236 & \pmark & 1 & 0.429\\
NF-TED$^\dagger$ (Syntactic) & 0.806 & 0.401 & 0.135 & \cmark & 3 & 0.463\\
Rule-profile (Syntactic) & 0.758 & 0.523 & 0.129 & \cmark & 3 & 0.432\\
\addlinespace
Vocabulary (Structural) & 0.933 & 0.498 & 0.261 & \pmark & 2 & 0.685\\
Typeclass-footprint (Structural) & 1.000 & 0.833 & 0.333 & \pmark & 1 & 0.833\\
Binder-profile (Structural) & 0.642 & 0.221 & 0.191 & \pmark & 2 & 0.595\\
\addlinespace
Direct-premises (Dependency \& Resource) & 0.982 & 0.271 & 0.123 & \pmark & 2 & 0.457\\
Dependency-Jaccard (Dependency \& Resource) & 1.000 & 0.656 & 0.447 & \cmark & 3 & 0.782\\
Dependency-GED (Dependency \& Resource) & 1.000 & 0.747 & 0.524 & \cmark & 3 & 0.852\\
Axiom-profile (Dependency \& Resource) & 1.000 & 1.000 & 1.000 & \xmark & 0 & 1.000\\
Module-EMD (Dependency \& Resource) & 0.990 & 0.776 & 0.702 & \pmark & 2 & 0.828\\
Size-ratio (Dependency \& Resource) & 0.758 & 0.561 & 0.143 & \pmark & 2 & 0.437\\
\addlinespace
Kernel-defeq (Semantic) & 1.000 & 1.000 & 1.000 & \xmark & 0 & 1.000\\
\addlinespace
NCD-source (Information-theoretic) & 0.434 & 0.274 & 0.246 & \pmark & 2 & 0.541\\
NCD-term (Information-theoretic) & 0.556 & 0.247 & 0.109 & \cmark & 3 & 0.351\\
\addlinespace
Shortest 2-cell (Higher-categorical) & 0.000 & 0.000 & 0.000 & \xmark & 0 & 0.000\\
Identification-level (Higher-categorical) & 0.333 & 0.000 & 0.000 & \xmark & 0 & 0.000\\
\bottomrule
\end{tabular}

%% file: lean_results/lean_properties.tex
\begin{tabular}{lccccccccccc}
\toprule
Metric (Group) & P1 & P2 & P3 & P4 & P5 & P6 & P7 & P8 & P9 & P10 & score\\
\midrule
Source-edit (Syntactic) & \cmark & \cmark & \cmark & \xmark & \xmark & \cmark & \cmark & \pmark & \cmark & \cmark & 7.5\\
Term-TED (Syntactic) & \cmark & \cmark & \cmark & \cmark & \xmark & \pmark & \cmark & \xmark & \pmark & \xmark & 6\\
NF-TED (Syntactic) & \cmark & \cmark & \cmark & \cmark & \xmark & \pmark & \cmark & \cmark & \pmark & \xmark & 7\\
Rule-profile (Syntactic) & \cmark & \cmark & \cmark & \cmark & \pmark & \xmark & \cmark & \pmark & \pmark & \xmark & 6.5\\
\addlinespace
Vocabulary (Structural) & \cmark & \cmark & \cmark & \cmark & \pmark & \pmark & \cmark & \pmark & \cmark & \pmark & 8\\
Typeclass-footprint (Structural) & \cmark & \cmark & \cmark & \cmark & \cmark & \xmark & \xmark & \xmark & \cmark & \pmark & 6.5\\
Binder-profile (Structural) & \cmark & \cmark & \cmark & \cmark & \xmark & \pmark & \cmark & \pmark & \cmark & \xmark & 7\\
\addlinespace
Direct-premises (Dependency \& Resource) & \cmark & \cmark & \cmark & \cmark & \pmark & \pmark & \cmark & \pmark & \cmark & \pmark & 8\\
Dependency-Jaccard (Dependency \& Resource) & \cmark & \cmark & \cmark & \cmark & \cmark & \pmark & \cmark & \cmark & \cmark & \pmark & 9\\
Dependency-GED (Dependency \& Resource) & \cmark & \cmark & \cmark & \cmark & \cmark & \pmark & \cmark & \cmark & \cmark & \pmark & 9\\
Axiom-profile (Dependency \& Resource) & \cmark & \cmark & \cmark & \cmark & \cmark & \xmark & \xmark & \xmark & \cmark & \xmark & 6\\
Module-EMD (Dependency \& Resource) & \cmark & \cmark & \cmark & \cmark & \cmark & \pmark & \cmark & \pmark & \cmark & \pmark & 8.5\\
Size-ratio (Dependency \& Resource) & \cmark & \cmark & \cmark & \cmark & \pmark & \xmark & \cmark & \pmark & \pmark & \pmark & 7\\
\addlinespace
Kernel-defeq (Semantic) & \cmark & \cmark & \cmark & \cmark & \cmark & \xmark & \xmark & \xmark & \cmark & \xmark & 6\\
\addlinespace
NCD-source (Information-theoretic) & \xmark & \xmark & \cmark & \xmark & \xmark & \cmark & \cmark & \pmark & \cmark & \cmark & 5.5\\
NCD-term (Information-theoretic) & \xmark & \xmark & \cmark & \xmark & \xmark & \cmark & \cmark & \cmark & \cmark & \xmark & 5\\
\addlinespace
Shortest 2-cell (Higher-categorical) & \cmark & \cmark & \cmark & \cmark & \xmark & \xmark & \cmark & \xmark & \pmark & \xmark & 5.5\\
Identification-level (Higher-categorical) & \cmark & \cmark & \cmark & \cmark & \xmark & \xmark & \cmark & \xmark & \pmark & \xmark & 5.5\\
\bottomrule
\end{tabular}

%% file: lean_results/lean_posthoc_module_idf.tex
\begin{tabular}{lllccc}
\toprule
Theorem & Proof & role & Module-EMD & IDF-weighted & closure-based\\
\midrule
primes & E$'$ & same & 0.989 & 1.000 & 1.000\\
primes & E$^{+}$ & finer & 0.940 & 0.960 & 0.843\\
primes & M & similar & 0.844 & 0.720 & 0.865\\
primes & G & similar & 0.792 & 0.828 & 0.959\\
primes & F & dissimilar & 0.608 & 0.818 & 0.907\\
primes & Eu & dissimilar & 0.532 & 0.639 & 0.840\\
sqrt2 & P$'$ & same & 0.997 & 1.000 & 1.000\\
sqrt2 & P$^{+}$ & finer & 0.840 & 0.879 & 0.950\\
sqrt2 & T & similar & 0.945 & 0.937 & 0.977\\
sqrt2 & V & similar & 0.706 & 0.851 & 0.889\\
sqrt2 & R & dissimilar & 0.759 & 0.824 & 0.886\\
gauss & I$'$ & same & 0.984 & 0.961 & 1.000\\
gauss & I$^{+}$ & finer & 0.704 & 0.900 & 0.882\\
gauss & Gp & similar & 0.683 & 0.882 & 0.839\\
gauss & H & dissimilar & 0.701 & 0.783 & 0.743\\
gauss & B & dissimilar & 0.850 & 0.843 & 0.802\\
\bottomrule
\end{tabular}

%% file: lean_results/lean_spearman_roles.tex
\begin{tabular}{lccccccc}
\toprule
Measure & Primes & $\sqrt2$ & Gauss & Mean & Pooled & $p$ & Pooled (no finer)\\
\midrule
Source-edit & $\mathbf{+0.96}$ & $+0.79$ & $+0.00$ & $+0.58$ & $+0.54$ & 0.016 & $+0.25$\\
Term-TED & $+0.72$ & $+0.37$ & $+0.63$ & $+0.57$ & $+0.29$ & 0.138 & $+0.29$\\
NF-TED & $\mathbf{+0.96}$ & $+0.74$ & $\mathbf{+0.79}$ & $+0.83$ & $+0.76$ & $<0.001$ & $+0.82$\\
Rule-profile & $+0.72$ & $+0.53$ & $+0.47$ & $+0.57$ & $+0.59$ & 0.008 & $+0.76$\\
Vocabulary & $\mathbf{+0.96}$ & $+0.74$ & $+0.63$ & $+0.78$ & $\mathbf{+0.81}$ & $<0.001$ & $\mathbf{+0.85}$\\
Typeclass-footprint & $+0.55$ & $+0.58$ & $+0.56$ & $+0.56$ & $+0.44$ & 0.050 & $+0.52$\\
Binder-profile & $+0.84$ & $+0.74$ & $+0.47$ & $+0.68$ & $+0.64$ & 0.005 & $+0.57$\\
Direct-premises & $\mathbf{+0.96}$ & $+0.74$ & $+0.63$ & $+0.78$ & $+0.76$ & $<0.001$ & $+0.76$\\
Dependency-Jaccard & $+0.72$ & $+0.95$ & $\mathbf{+0.79}$ & $+0.82$ & $+0.69$ & 0.002 & $+0.71$\\
Dependency-GED & $+0.48$ & $+0.95$ & $\mathbf{+0.79}$ & $+0.74$ & $+0.64$ & 0.005 & $+0.71$\\
Axiom-profile & -- & -- & -- & -- & -- & -- & --\\
Module-EMD & $\mathbf{+0.96}$ & $+0.58$ & $+0.32$ & $+0.62$ & $+0.62$ & 0.006 & $+0.73$\\
Size-ratio & $+0.72$ & $+0.53$ & $+0.32$ & $+0.52$ & $+0.56$ & 0.014 & $+0.73$\\
Kernel-defeq & -- & -- & -- & -- & -- & -- & --\\
NCD-source & $+0.91$ & $+0.79$ & $+0.32$ & $+0.67$ & $+0.66$ & 0.003 & $+0.45$\\
NCD-term & $\mathbf{+0.96}$ & $\mathbf{+0.97}$ & $\mathbf{+0.79}$ & $\mathbf{+0.91}$ & $+0.52$ & 0.020 & $+0.58$\\
Shortest 2-cell & -- & -- & -- & -- & -- & -- & --\\
Identification-level & $+0.55$ & $+0.56$ & -- & -- & $+0.43$ & 0.123 & $+0.61$\\
\bottomrule
\end{tabular}

%% file: lean_results/lean_independence.tex
\begin{tabular}{llllc}
\toprule
Theorem & Proof & role & library equivalent in closure & chain length\\
\midrule
primes & E & base & -- & --\\
primes & E$'$ & same & -- & --\\
primes & E$^{+}$ & finer & -- & --\\
primes & M & similar & \texttt{Nat.exists\_infinite\_primes} & 0\\
primes & G & similar & -- & --\\
primes & F & dissimilar & -- & --\\
primes & Eu & dissimilar & \texttt{Nat.exists\_infinite\_primes} & 9\\
sqrt2 & P & base & -- & --\\
sqrt2 & P$'$ & same & -- & --\\
sqrt2 & P$^{+}$ & finer & -- & --\\
sqrt2 & T & similar & -- & --\\
sqrt2 & V & similar & -- & --\\
sqrt2 & R & dissimilar & \texttt{irrational\_sqrt\_two} & 0\\
gauss & I & base & -- & --\\
gauss & I$'$ & same & -- & --\\
gauss & I$^{+}$ & finer & -- & --\\
gauss & Gp & similar & -- & --\\
gauss & H & dissimilar & -- & --\\
gauss & B & dissimilar & -- & --\\
\bottomrule
\end{tabular}

%% file: lean_results/lean_cross_theorem.tex
\begin{tabular}{lccll}
\toprule
Measure & within & across & most similar cross-theorem pair & $s$\\
\midrule
Dependency-Jaccard & 0.58 & 0.50 & M (Primes) -- I$^{+}$ (Gauss) & 0.76\\
Direct-premises & 0.26 & 0.11 & T ($\sqrt2$) -- B (Gauss) & 0.37\\
Vocabulary & 0.43 & 0.22 & G (Primes) -- Gp (Gauss) & 0.43\\
Rule-profile & 0.33 & 0.22 & E$^{+}$ (Primes) -- Gp (Gauss) & 0.85\\
Binder-profile & 0.28 & 0.17 & E$'$ (Primes) -- V ($\sqrt2$) & 0.81\\
Module-EMD & 0.75 & 0.69 & G (Primes) -- Gp (Gauss) & 0.88\\
\bottomrule
\end{tabular}

%% file: applications.tex
\section{What a metric over proofs could enable}\label{sec:enable}
A measure of proof dissimilarity is only worth having if it changes what we can do.
We see four uses (\S\ref{sec:enable}.1--\ref{sec:enable}.4), illustrate them with the 2026 announcements (\S\ref{sec:events}), and end with one danger; throughout, we map the uses onto the properties of \S\ref{sec:props}.
Table~\ref{tab:uses} summarises the mapping.

\subsection{Checking and evaluating machine-generated proofs}
\paragraph{Formal proofs.}
When an LLM produces a Lean proof, the kernel settles \emph{correctness}. What it cannot settle is what the proof \emph{is}:
\begin{itemize}[leftmargin=*,itemsep=1pt]
\item whether it recalls a known proof (contamination or memorisation) or finds a new one;
\item whether $k$ samples are $k$ proofs or one proof written $k$ ways;
\item whether reinforcement learning has taught a prover a new strategy, or only a heavier use of decision procedures.
\end{itemize}
Nearest-neighbour search under a bureaucracy-invariant measure (P4, P5), against a library and a training corpus, answers the first question. \textsf{Dependency-Jaccard} is the natural choice: it satisfied P5 in both studies, whereas \textsf{NF-TED} did so only in the kernel study.
Diversity of a sample under the same measures answers the second. It suggests reporting \emph{distinct-proof@k} next to pass@$k$.
Profiles of \textsf{Direct-premises} and closure-based area measures over training time answer the third. (\textsf{Axiom-profile} is not useful here, because it is constant across Mathlib; \S\ref{sec:lean}.)

\paragraph{Interpreting machine-generated proofs.}
A proof that is correct but opaque is of limited use to mathematicians. A dissimilarity measure supports a simple interpretability procedure.
Retrieve the \emph{most similar human-understandable proofs}, of the same theorem or of related ones, from a library or from the literature, and align them with the generated proof at the levels where they agree (shared dependencies, construction, computational content).
Agreement lets the familiar argument explain the unfamiliar one. Divergence localises what is new and where human attention is needed.

\paragraph{Orchestrating prover agents.}
Beyond merely evaluating proof search, similarity can also be used to steer it.
In multi-agent provers, agents can be assigned and redirected so that their partial \emph{proof trajectories} (the sequences of intermediate lemmas, tactics and dependencies they accumulate) remain mutually dissimilar.
This maximises the complementarity of the agents and reduces the effort wasted when several of them independently rediscover the same route.
Concretely, a coordinator can penalise new work in proportion to its similarity to trajectories already in progress, or allocate agents to under-explored regions of proof space, as in quality-diversity search~\cite{mapelites}. Bureaucracy and granularity invariance (P5, P6) ensure that agents are not rewarded merely for rephrasing each other.

\paragraph{Informal proofs.}
For natural-language proofs there is no kernel, and fluent wrong proofs are common~\cite{proofbluff}.
A similarity measure cannot certify correctness: a one-symbol edit can turn a valid argument into an invalid one, so validity is not continuous in any syntactic metric.
It can, however, \emph{triage}.
Align a generated proof with its nearest known proof, at the level of dependencies, conceptual vocabulary and step structure. Where the alignment holds, the generated proof inherits plausibility. Where it breaks (an unfamiliar lemma, a step with no counterpart, an abrupt jump in \textsf{Module-EMD}), human or autoformalisation effort is concentrated.
The same alignment is a check of \emph{autoformalisation fidelity}. A formalisation can prove the right statement by a different argument, for instance by calling a library theorem or a decision procedure, and still be marked correct. Comparing the structure of the informal proof with that of the formal one detects this, and statement-level checks cannot.

\subsection{Understanding proofs conceptually}
\paragraph{Families of proofs.}
For theorems with many proofs, clustering under a profile of measures turns informal taxonomies, such as Lemmermeyer's grouping of reciprocity proofs by method~\cite{lemmermeyer}, into testable claims.
Disagreement between levels is informative. Our kernel study found two proofs that are far apart in concepts but identical in computational content (Furstenberg and Euclid). Systematic search for such pairs is a search for \emph{proofs in disguise}, which matters for the history of mathematics and for transferring techniques between fields.

\paragraph{Explanation, purity and simplicity.}
Philosophers have argued about explanatory and pure proofs mostly through examples~\cite{purity,steiner,lange}.
Measures such as \textsf{Module-EMD} (distance between the subject areas a proof draws on and the area of its statement), \textsf{Hypothesis-usage} (generality) and, once it can locate \emph{essential} uses of classical axioms, \textsf{Axiom-profile} are candidate correlates.
Combined with appraisal data of the kind collected by Inglis and Aberdein~\cite{inglis}, they make such philosophical hypotheses empirically testable.
Hilbert's original question about \emph{simplicity} becomes, in this setting, a question about which measure, if any, tracks expert judgements of simplicity.

\paragraph{Analogy and generalisation across theorems.}
Our measures compare proofs of one theorem. Several of them, for example \textsf{Flow-linking} up to renaming of lemmas, \textsf{Rule-profile} and \textsf{Binder-profile}, also compare proofs of \emph{different} theorems.
Proofs of different theorems that are close in such a measure share a skeleton.
Anti-unifying them~\cite{plotkin,reynolds} proposes the common generalisation, in effect a new lemma, as library-learning systems do for programs~\cite{dreamcoder}.
This makes Pólya's heuristic ``do you know a related problem?'' operational.

\subsection{Mapping proof space}
With a dissimilarity measure, the proofs of a theorem form a (pseudo)metric space, and the question that opened this project has a precise form: \emph{how large is its diameter?}
Several notions follow.
\begin{itemize}[leftmargin=*,itemsep=1pt]
\item \emph{Diameter and coverage}: how far apart the known proofs are, and how much of the space they cover.
\item \emph{Clusters and gaps}: families of proofs, and regions between them with no known proof.
\item \emph{Strata}: the ultrametric of \S\ref{sec:metrics} organises proofs into a tree (same term, same normal form, same generality, same content, same theorem). A theorem's proofs can be displayed as that tree, with metric detail inside each node.
\item \emph{Embeddings}: multidimensional scaling or topological data analysis of distance matrices like Appendix~\ref{app:allpairs} give pictures of proof space.
\end{itemize}
Two caveats are essential.
First, the raw space is trivially unbounded: any proof can be padded with detours. Diameter is meaningful only on proofs quotiented by bureaucracy (normal forms, or minimal proofs), which is exactly what P5 demands.
Second, the shape of the space depends on the library and the granularity of the encoding (P6). Proof space is always \emph{relative to a library}, and a growing library changes it.

\paragraph{Exploring proof space on purpose.}
A metric can also be used generatively.
Quality-diversity search~\cite{mapelites} with a prover as generator, correctness as the quality criterion and a proof measure as the novelty criterion would look for maximally different proofs of a theorem. Producing new proofs of old theorems is a recognised mathematical activity~\cite{dawson}.
Conditioning on a stratum, for example ``a proof with the same Herbrand content but in a different area'', or ``a proof using no \texttt{Classical.choice}'', turns proof search into directed exploration of the strata above.

\subsection{Engineering formal libraries}
In large libraries, proof similarity supports routine tasks:
\begin{itemize}[leftmargin=*,itemsep=1pt]
\item detecting duplicated arguments and candidates for factoring out a lemma;
\item choosing which existing proof to adapt during proof repair~\cite{qed};
\item ranking premises~\cite{alama,kaliszykurban};
\item measuring the effect of a refactoring on how proofs depend on each other (\textsf{Dependency-GED}).
\end{itemize}

\subsection{The 2026 announcements as test cases}\label{sec:events}
The episodes recounted in the introduction make the uses above concrete, and suggest several that we had not anticipated.
\begin{description}[leftmargin=1.2em,itemsep=3pt]
\item[Priority and independence.]
The Navier--Stokes/Euler dispute~\cite{oai-ns,quanta-ns,techcrunch-ns} turns on whether two arguments are ``the same''.
Both sides say the strategy goes back to C\'ordoba and Mart\'{\i}nez-Zoroa. OpenAI says its proof differs, and in particular that its Euler result is for the unforced equations.
Our framework separates the questions being conflated:
\begin{itemize}[leftmargin=*,itemsep=0pt]
\item the statements differ (forced vs unforced), so this is \emph{cross-theorem} similarity;
\item the arguments may agree at the level of construction and computational content (the blow-up profile) while differing in lemmas and presentation;
\item both arguments exist as Lean formalisations, so the exporter of \S\ref{sec:lean} applies to them directly.
\end{itemize}
A profile of distances would not settle priority, which is a historical question. It would replace ``it is obviously the same / obviously different'' with a statement of \emph{at which levels} the arguments coincide.
\item[Contamination and data provenance.]
Whether a user's interactions with an AI product could have informed a model's later proof~\cite{techcrunch-ns} is the contamination question of \S\ref{sec:enable} in its most consequential form.
Proof similarity can give evidence, but only one-sided evidence. High similarity at every level is consistent with both copying and convergent discovery from a shared source (here, the C\'ordoba--Mart\'{\i}nez-Zoroa strategy). Low similarity at the levels that matter is evidence of independence.
Any forensic use must account for this asymmetry, and for the fact that the relevant inputs (prompts, private drafts) are rarely available.
\item[Quantifying human versus machine contribution.]
Human involvement in these results ranged from setting goals and steering agents to supplying the key idea.
The directed quantity $\hat K(B\mid A)$ of \S\ref{sec:metrics}, how much information the final proof $B$ contains beyond the human inputs $A$ (problem formulation, hints, sketches, earlier papers), is a candidate for measuring that contribution.
It is crude, but it is the right \emph{kind} of quantity: asymmetric, and measured relative to what the humans actually supplied.
\item[Swarms and consolidation.]
A run with $10^4$ agents and millions of messages~\cite{oai-ns,quanta-ns} faces our questions internally.
Deciding which partial arguments are duplicates, which are genuinely different routes, and which should be merged requires a measure of proof dissimilarity. So does orchestrating the agents so that their proof trajectories are complementary rather than redundant (\S\ref{sec:enable}), which reduces the wasted effort of thousands of agents converging on the same route. Bureaucracy and granularity invariance (P5, P6) are what stop agents being rewarded for rephrasing.
\item[Families and corpora.]
The 721 manuscripts in 373 families~\cite{oai-math} are already organised by an implicit similarity judgement.
They are also, with about 42\% of top-line results formalised, the first large corpus of machine-generated research mathematics.
Our measures could audit the family structure, place each family relative to the human literature (nearest prior work, novelty, area distance), and triage the unformalised results for review by their distance from formalised neighbours.
\item[Computational content of disproofs.]
A disproof of the exponential time hypothesis~\cite{oai-oct} asserts that 3-SAT has subexponential algorithms. Whether it \emph{yields} one depends on whether the proof is constructive, which is the semantic level of our framework.
Our Lean study shows that \texttt{\#print axioms}, and hence \textsf{Axiom-profile}, cannot answer this: every Mathlib proof we examined depends on \texttt{Classical.choice}. What is needed is an analysis of \emph{where} choice is used, and whether it is essential. If the proof turns out to be constructive, proof-mining techniques~\cite{kohlenbach} could extract the algorithm.
Two independently found disproofs would then be compared by whether they encode \emph{the same algorithm}. That is the Herbrand-level identity under which our kernel study found Furstenberg's proof to be Euclid's.
\item[Proof by a stronger theorem.]
The Kakeya and restriction conjectures were obtained as consequences of local smoothing~\cite{oai-oct}.
In our terms these are ``citation proofs'', like M, Eu and R in \S\ref{sec:lean}: short at the surface, with an enormous dependency closure. They are exactly the cases where surface measures and dependency measures disagree most.
\item[Comprehensibility.]
When experts find a machine-generated proof hard to read~\cite{quanta-ns}, part of the difficulty is distance from familiar arguments.
Distance to the nearest human proof, \textsf{Module-EMD} to the theorem's own area, and size of the normal form are candidate, testable proxies for comprehensibility. They could flag proofs that sit in regions of proof space no human proof occupies. Retrieving the nearest human-understandable proof is also a constructive aid: it gives readers a familiar argument to read the machine proof against.
\item[Statement fidelity.]
Reports stress that a Lean check certifies the proof of the \emph{formal} statement, and that humans must still confirm that this statement is the intended one~\cite{quanta-ns}.
That is a different, sibling problem (similarity of \emph{statements}). Proof measures help only indirectly, for instance by revealing that a formal proof's dependencies or area profile are implausible for the intended theorem.
\end{description}

\subsection{The danger: Goodhart's law}
As soon as a measure is used as an objective, for example as a novelty reward for a prover, it will be gamed.
A measure that fails $\alpha$-invariance (P4) rewards renaming. One that fails bureaucracy invariance (P5) rewards detours. One that fails granularity robustness (P6) rewards splitting lemmas. One that relies only on vocabulary rewards replacing a lemma by an equivalent one from another library area.
The properties of \S\ref{sec:props} are therefore not academic: they are the conditions under which a proof measure can safely be \emph{optimised}.
On our results, no single measure is safe on its own (Tables~\ref{tab:props} and~\ref{tab:leanprops}). This is a further argument for profiles. A proof should count as novel only if it is novel at several levels at once.

\begin{table}[t]
\centering\small
\begin{tabular}{p{3.6cm}p{3.1cm}p{6.6cm}}
\toprule
Use & Properties needed & Candidate measures (kernel study / Lean study)\\
\midrule
Contamination / recall detection & P4, P5, P9 & Dependency-Jaccard/GED, Direct-premises; NF-TED within normal-form classes\\
Diversity of generated proofs (distinct-proof@$k$) & P4, P5, P6, P7 & Dependency-GED, Module-EMD, Flow-linking; Identification-level as a coarse filter\\
Orchestrating prover agents for complementarity & P4--P7 & dependency and lemma-trajectory measures (Dependency-GED, Direct-premises); a profile for robustness\\
Interpreting AI proofs via nearest human proofs & P5, P7, P8 & Dependency-Jaccard, NF-TED; Module-EMD for area context\\
Triage of informal LLM proofs & P10, P7 & NCD-text (weak); MSC-/Module-EMD and dependency alignment after extracting cited results\\
Autoformalisation fidelity & P5, P7, P10 & dependency and vocabulary alignment between informal and formal proof\\
Clustering proof families & P1--P3, P6, P8 & Dependency-Jaccard/GED, MSC-/Module-EMD; semantic measures to find proofs in disguise\\
Purity / explanation studies & P8, P10 & MSC-/closure-based Module-EMD, Hypothesis-usage; Axiom-profile only with a finer analysis of choice\\
Cross-theorem analogy & P4, P5 & Flow-linking (up to lemma renaming), Rule-profile, Binder-profile\\
Novelty-driven proof search & P4--P7 (all) & a profile; no single measure suffices\\
Library maintenance, proof repair & P5, P9 & Direct-premises, Dependency-GED, NF-TED\\
Independence / relative-circularity check & P9 & dependency closure with library equivalents of the claim (Table~\ref{tab:indep})\\
\bottomrule
\end{tabular}
\caption{Uses of a proof-dissimilarity measure, the properties each use requires, and the measures from this paper that currently come closest. Recommendations combine both studies (Tables~\ref{tab:props} and~\ref{tab:leanprops}); measures that are weak or degenerate in Lean are discussed in \S\ref{sec:discussion}.}
\label{tab:uses}
\end{table}

%% file: toy_results/library.tex
\begin{tabular}{l p{6.2cm} l l}
\toprule
Lemma & Content & MSC & used by\\
\midrule
\texttt{prod\_ge2} & prod(S)+1 >= 2 for finite S & 11A25 & E, E$'$, E$^{+}$\\
\texttt{lpf\_prime} & the least factor >1 of N>=2 is prime & 11A51 & E, E$'$, E$^{+}$, G, F\\
\texttt{lpf\_dvd} & lpf(N) divides N & 11A51 & E, E$'$, E$^{+}$, G, F\\
\texttt{dvd\_prod} & a member of S divides prod(S) & 11A05 & E, E$'$\\
\texttt{prime\_dvd\_succ} & a prime cannot divide both a and a+1 & 11A41 & E, E$'$, E$^{+}$\\
\texttt{mem\_split} & prod(S) = q * prod(S minus q) for q in S & 11A25 & E$^{+}$\\
\texttt{dvd\_mul} & a | a*b & 11A05 & E$^{+}$\\
\texttt{dvd\_eq\_r} & rewrite the dividend along an equation & 03B10 & E$^{+}$\\
\texttt{F\_ge2} & Fermat numbers are >= 2 & 11A51 & G\\
\texttt{pigeon} & pigeonhole: |S|+1 values in S collide & 05D10 & G\\
\texttt{fermat\_coprime} & distinct Fermat numbers are coprime & 11A05 & G\\
\texttt{dvd\_eq} & rewrite the divisor along an equation & 03B10 & G\\
\texttt{mult\_closed} & pZ is closed (its complement is a union of progressions) & 54A05, 11B25 & F\\
\texttt{union\_closed} & finite unions of closed sets are closed & 54A05 & F\\
\texttt{closed\_open} & complements of closed sets are open & 54A05 & F\\
\texttt{one\_comp} & 1 is not a multiple of any prime & 11A41 & F\\
\texttt{open\_unbounded} & a nonempty open set contains arbitrarily large elements & 54A05, 11B25 & F\\
\texttt{gt1\_ge2} & y > 1 implies y >= 2 & 03F30 & F\\
\texttt{mem\_umult} & a multiple of some p in S lies in the union & 11A05, 03E20 & F\\
\texttt{comp\_contra} & nothing lies in a set and its complement & 03E20 & F\\
\bottomrule
\end{tabular}

%% file: toy_results/distance_all_pairs.tex
\begin{tabular}{lcccccccccc}
\toprule
Metric & E,E$'$ & E,E$^{+}$ & E,G & E,F & E$'$,E$^{+}$ & E$'$,G & E$'$,F & E$^{+}$,G & E$^{+}$,F & G,F\\
\midrule
Raw TED & 0.49 & 0.24 & 0.50 & 0.49 & 0.53 & 0.58 & 0.61 & 0.53 & 0.50 & 0.47\\
$\alpha$-TED & 0.44 & 0.24 & 0.50 & 0.49 & 0.49 & 0.58 & 0.57 & 0.53 & 0.50 & 0.45\\
NF-TED & 0.00 & 0.22 & 0.49 & 0.45 & 0.22 & 0.49 & 0.45 & 0.55 & 0.49 & 0.48\\
Rule-profile & 0.00 & 0.17 & 0.27 & 0.36 & 0.17 & 0.27 & 0.36 & 0.15 & 0.23 & 0.21\\
Flow-linking & 0.00 & 0.57 & 0.88 & 0.92 & 0.57 & 0.88 & 0.92 & 0.88 & 0.92 & 0.91\\
Hypothesis-usage & 0.00 & 0.00 & 0.00 & 0.33 & 0.00 & 0.00 & 0.33 & 0.00 & 0.33 & 0.33\\
Signature & 0.00 & 0.27 & 0.54 & 0.57 & 0.27 & 0.54 & 0.57 & 0.53 & 0.65 & 0.65\\
Dependency-Jaccard & 0.00 & 0.25 & 0.57 & 0.67 & 0.25 & 0.57 & 0.67 & 0.61 & 0.69 & 0.71\\
Dependency-GED & 0.00 & 0.16 & 0.50 & 0.60 & 0.16 & 0.50 & 0.60 & 0.53 & 0.61 & 0.64\\
Axiomatic-strength & 0.00 & 0.00 & 0.00 & 0.67 & 0.00 & 0.00 & 0.67 & 0.00 & 0.67 & 0.67\\
MSC-EMD & 0.00 & 0.16 & 0.35 & 0.59 & 0.16 & 0.35 & 0.59 & 0.22 & 0.51 & 0.51\\
Size-ratio & 0.00 & 0.36 & 0.09 & 0.17 & 0.36 & 0.09 & 0.17 & 0.29 & 0.22 & 0.09\\
Herbrand-witness & 0.00 & 0.00 & 0.84 & 0.00 & 0.00 & 0.84 & 0.00 & 0.84 & 0.00 & 0.84\\
Query-trace & 0.00 & 0.00 & 0.86 & 0.00 & 0.00 & 0.86 & 0.00 & 0.86 & 0.00 & 0.86\\
Witness-magnitude & 0.00 & 0.00 & 0.47 & 0.00 & 0.00 & 0.47 & 0.00 & 0.47 & 0.00 & 0.47\\
NCD-term & 0.07 & 0.27 & 0.57 & 0.65 & 0.27 & 0.57 & 0.65 & 0.56 & 0.64 & 0.62\\
NCD-text & 0.69 & 0.17 & 0.70 & 0.73 & 0.72 & 0.80 & 0.81 & 0.70 & 0.73 & 0.76\\
Shortest 2-cell & 0.93 & 1.00 & 1.00 & 1.00 & 1.00 & 1.00 & 1.00 & 1.00 & 1.00 & 1.00\\
Identification-level & 0.25 & 0.75 & 1.00 & 0.75 & 0.75 & 1.00 & 0.75 & 1.00 & 0.75 & 1.00\\
\bottomrule
\end{tabular}

%% file: lean_results/lean_similarity_primes.tex
\begin{tabular}{lcccccc}
\toprule
Metric & E' & E+ & M & G & F & Eu\\
 & \emph{same} & \emph{finer} & \emph{similar} & \emph{similar} & \emph{dissimilar} & \emph{dissimilar}\\
\midrule
Source-edit & 0.34 & 0.59 & 0.18 & 0.22 & 0.10 & 0.14\\
Term-TED & 0.46 & 0.69 & 0.43 & 0.47 & 0.43 & 0.39\\
NF-TED & 0.90 & 0.64 & 0.46 & 0.62 & 0.30 & 0.04\\
Rule-profile & 0.95 & 0.47 & 0.53 & 0.79 & 0.35 & 0.02\\
Vocabulary & 0.92 & 0.82 & 0.73 & 0.57 & 0.20 & 0.16\\
Typeclass-footprint & 1.00 & 1.00 & 1.00 & 1.00 & 0.00 & 1.00\\
Binder-profile & 0.93 & 0.29 & 0.06 & 0.17 & 0.02 & 0.08\\
Direct-premises & 1.00 & 0.69 & 0.39 & 0.36 & 0.09 & 0.03\\
Dependency-Jaccard & 1.00 & 0.71 & 0.66 & 0.88 & 0.70 & 0.33\\
Dependency-GED & 1.00 & 0.82 & 0.78 & 0.92 & 0.82 & 0.40\\
Axiom-profile & 1.00 & 1.00 & 1.00 & 1.00 & 1.00 & 1.00\\
Module-EMD & 0.99 & 0.94 & 0.84 & 0.79 & 0.61 & 0.53\\
Size-ratio & 0.95 & 0.47 & 0.53 & 0.79 & 0.43 & 0.02\\
Kernel-defeq & 1.00 & 1.00 & 1.00 & 1.00 & 1.00 & 1.00\\
NCD-source & 0.50 & 0.65 & 0.30 & 0.32 & 0.10 & 0.30\\
NCD-term & 0.82 & 0.62 & 0.54 & 0.55 & 0.17 & 0.10\\
Shortest 2-cell & 0.00 & 0.00 & 0.00 & 0.00 & 0.00 & 0.00\\
Identification-level & 0.50 & 0.00 & 0.00 & 0.00 & 0.00 & 0.00\\
\bottomrule
\end{tabular}

%% file: lean_results/lean_similarity_sqrt2.tex
\begin{tabular}{lccccc}
\toprule
Metric & P' & P+ & T & V & R\\
 & \emph{same} & \emph{finer} & \emph{similar} & \emph{similar} & \emph{dissimilar}\\
\midrule
Source-edit & 0.34 & 0.37 & 0.28 & 0.14 & 0.23\\
Term-TED & 0.04 & 0.26 & 0.65 & 0.02 & 0.03\\
NF-TED & 0.82 & 0.42 & 0.56 & 0.03 & 0.02\\
Rule-profile & 0.75 & 0.59 & 0.78 & 0.03 & 0.01\\
Vocabulary & 0.98 & 0.61 & 0.85 & 0.37 & 0.23\\
Typeclass-footprint & 1.00 & 0.50 & 1.00 & 0.00 & 0.00\\
Binder-profile & 0.65 & 0.57 & 0.59 & 0.12 & 0.03\\
Direct-premises & 1.00 & 0.38 & 0.61 & 0.16 & 0.08\\
Dependency-Jaccard & 1.00 & 0.69 & 0.67 & 0.58 & 0.33\\
Dependency-GED & 1.00 & 0.76 & 0.74 & 0.66 & 0.38\\
Axiom-profile & 1.00 & 1.00 & 1.00 & 1.00 & 1.00\\
Module-EMD & 1.00 & 0.84 & 0.95 & 0.71 & 0.76\\
Size-ratio & 0.75 & 0.61 & 0.79 & 0.03 & 0.01\\
Kernel-defeq & 1.00 & 1.00 & 1.00 & 1.00 & 1.00\\
NCD-source & 0.52 & 0.38 & 0.36 & 0.23 & 0.24\\
NCD-term & 0.20 & 0.08 & 0.02 & 0.02 & 0.02\\
Shortest 2-cell & 0.00 & 0.00 & 0.00 & 0.00 & 0.00\\
Identification-level & 0.50 & 0.00 & 0.00 & 0.00 & 0.00\\
\bottomrule
\end{tabular}

%% file: lean_results/lean_similarity_gauss.tex
\begin{tabular}{lccccc}
\toprule
Metric & I' & I+ & Gp & H & B\\
 & \emph{same} & \emph{finer} & \emph{similar} & \emph{dissimilar} & \emph{dissimilar}\\
\midrule
Source-edit & 0.12 & 0.40 & 0.10 & 0.22 & 0.20\\
Term-TED & 0.94 & 0.34 & 0.24 & 0.28 & 0.25\\
NF-TED & 0.69 & 0.34 & 0.37 & 0.26 & 0.18\\
Rule-profile & 0.57 & 0.23 & 0.51 & 0.24 & 0.14\\
Vocabulary & 0.90 & 0.62 & 0.23 & 0.46 & 0.27\\
Typeclass-footprint & 1.00 & 1.00 & 1.00 & 1.00 & 0.00\\
Binder-profile & 0.34 & 0.93 & 0.19 & 0.84 & 0.13\\
Direct-premises & 0.95 & 0.30 & 0.05 & 0.20 & 0.27\\
Dependency-Jaccard & 1.00 & 0.94 & 0.57 & 0.75 & 0.24\\
Dependency-GED & 1.00 & 0.98 & 0.69 & 0.86 & 0.31\\
Axiom-profile & 1.00 & 1.00 & 1.00 & 1.00 & 1.00\\
Module-EMD & 0.98 & 0.70 & 0.68 & 0.70 & 0.85\\
Size-ratio & 0.57 & 0.23 & 0.62 & 0.24 & 0.14\\
Kernel-defeq & 1.00 & 1.00 & 1.00 & 1.00 & 1.00\\
NCD-source & 0.29 & 0.60 & 0.22 & 0.33 & 0.28\\
NCD-term & 0.65 & 0.35 & 0.17 & 0.24 & 0.11\\
Shortest 2-cell & 0.00 & 0.00 & 0.00 & 0.00 & 0.00\\
Identification-level & 0.00 & 0.00 & 0.00 & 0.00 & 0.00\\
\bottomrule
\end{tabular}

%% file: glossary.tex
\section{Glossary of non-standard terms}\label{app:glossary}
Terms marked $^\dagger$ are introduced in this paper or used here in a specific sense; the others are standard in logic, proof assistants or computer science, but may be unfamiliar to some readers.

\begin{description}[leftmargin=1em,itemsep=2pt,font=\normalfont\bfseries]
\item[$\alpha$-equivalence, $\alpha$-variant.] Two terms are $\alpha$-equivalent if they differ only in the names of bound variables. An $\alpha$-variant$^\dagger$ is a copy of a proof with every bound name changed, used to test property P4.
\item[Axioms (Lean): \texttt{propext}, \texttt{Classical.choice}, \texttt{Quot.sound}.] The three axioms Lean adds to its type theory: propositional extensionality, the axiom of choice (which makes the logic classical), and soundness of quotient types. \texttt{\#print axioms} lists those a proof depends on.
\item[Base proof$^\dagger$.] For each theorem, the proof against which the others are compared (E, P, I).
\item[Bureaucracy.] Inessential presentational differences between derivations, such as the order of independent steps, extracting a lemma, introducing and immediately using an abbreviation (\texttt{let}/\texttt{have}), or detours that normalisation removes. \emph{Bureaucracy invariance}$^\dagger$ (P5) is insensitivity to such differences.
\item[Citation proof$^\dagger$ (proof by citation).] A proof that obtains its result mainly by invoking a library theorem that already contains the argument (M, Eu, R, B; also the Kakeya result via local smoothing). Short at the surface, with a large dependency closure.
\item[Closure, dependency closure.] The set of all library constants a proof depends on, directly or transitively, through the types and values of the constants it uses.
\item[Curry-style tree.] A proof term with type annotations erased, keeping only its computational structure. Our Lean trees also erase implicit and type-class arguments.
\item[de Bruijn index.] A representation of bound variables by numbers (``the variable bound $k$ binders up'') instead of names. It makes $\alpha$-equivalent terms syntactically equal.
\item[Definitional equality; proof irrelevance.] Lean's kernel considers two terms definitionally equal if they reduce to a common form. By \emph{definitional proof irrelevance}, any two proofs of the same proposition are definitionally equal, regardless of their content.
\item[$\delta$, $\beta$, $\zeta$, $\iota$-reduction.] Unfolding a definition ($\delta$); applying a function to its argument ($\beta$); unfolding a local \texttt{let}/\texttt{have} ($\zeta$); simplifying a projection or match applied to a constructor ($\iota$). Our \emph{local normal form}$^\dagger$ applies $\delta$ only to the proof's own auxiliary constants, never to library constants.
\item[Distinct-proof@$k$$^\dagger$.] Proposed evaluation metric: the number of mutually dissimilar correct proofs among $k$ generated attempts, reported next to pass@$k$.
\item[Earth mover's distance (EMD).] The minimal cost of transporting one probability distribution onto another; on a tree it has a closed form. MSC-EMD and Module-EMD$^\dagger$ apply it to distributions over subject-classification codes and over Mathlib modules.
\item[Elaboration; elaborated term.] The translation of a user-written Lean proof (often a tactic script) into a kernel-checkable proof term. Two scripts that read alike may elaborate to very different terms.
\item[Euclid's witness.] The concrete prime produced by Euclid's argument when it is run against a purported finite list $S$ of all primes: the least prime factor of $1+\prod S$, which cannot belong to $S$. For example, for $S=\{2,3,5,7\}$ it is $211$. The kernel study's semantic measures compare proofs by the witnesses they produce, and Furstenberg's proof produces this same prime for every $S$.
\item[Exponential time hypothesis (ETH).] The conjecture that 3-SAT cannot be solved in time $2^{o(n)}$; a disproof implies that subexponential algorithms exist.
\item[Flow graph; generality.] In the Lambek--Do\v{s}en tradition, the pattern of links between occurrences of atomic formulas established by a proof. Two proofs with the same links have the same \emph{generality}. Our Flow-linking$^\dagger$ measure compares these links.
\item[Granularity robustness$^\dagger$ (P6).] Insensitivity to the level of detail at which a proof is written, e.g.\ one lemma versus three smaller lemmas.
\item[Herbrand witness; no-counterexample reading.] Running a refutation against a purported counterexample yields a concrete object that defeats it; here, the prime that a purported finite list of all primes misses. Our semantic measures$^\dagger$ compare proofs by these witnesses.
\item[I$\Delta_0$, RCA$_0$, ACA$_0$, Z$_2$.] Formal systems of arithmetic of increasing strength (bounded induction; recursive comprehension; arithmetical comprehension; full second-order arithmetic). Used in reverse mathematics to calibrate what a theorem or proof needs.
\item[Identification level$^\dagger$; ultrametric.] The coarsest level in our ladder of identification criteria (same term, same normal form, same premises, same computational content or area, same theorem) at which two proofs are identified. The resulting distance is an ultrametric: $d(x,z)\le\max(d(x,y),d(y,z))$.
\item[Independence; relative circularity$^\dagger$.] A proof is relatively circular if its dependency closure contains a library statement that trivially implies the theorem it proves. Two proofs are independent if neither depends on the other's key lemmas or on a library equivalent of the theorem.
\item[IDF weighting.] Inverse document frequency: down-weighting items that occur in many proofs (here, ubiquitous bookkeeping lemmas).
\item[Kernel.] Used in two senses. (i) The \emph{kernel of a proof assistant} is its small, trusted core, which checks every step of a formal proof; Lean's kernel certifies correctness but says nothing about what a proof contains or how it relates to other proofs. (ii) Our \emph{Curry--Howard kernel} is the minimal proof checker built for the first study (\S\ref{sec:case}), in which proofs are typed terms that can be type-checked, normalised and run.
\item[Lean, Mathlib, tactic, \texttt{sorry}.] Lean~4 is an interactive theorem prover; Mathlib is its community mathematics library. A tactic is a command that builds part of a proof automatically (e.g.\ \texttt{simp}, \texttt{omega}, \texttt{nlinarith}). \texttt{sorry} is a placeholder for a missing proof, recorded as an axiom.
\item[lpf.] Least prime factor ($\mathrm{lpf}(n)$; \texttt{Nat.minFac} in Mathlib).
\item[MinHash, locality-sensitive hashing (LSH).] Techniques for finding pairs of large sets with high Jaccard similarity without comparing all pairs.
\item[MSC.] The Mathematics Subject Classification (MSC2020) of the AMS and zbMATH.
\item[NCD.] Normalised compression distance: an approximation to the (uncomputable) information distance, obtained with a standard compressor.
\item[Normal form; normalisation; convertibility.] A normal form is a proof with no remaining reducible detours; normalisation computes it. Two proofs are convertible if they reduce to a common term. Prawitz's normalisation conjecture identifies proofs with equal normal forms.
\item[pass@$k$.] The probability that at least one of $k$ sampled attempts is correct.
\item[pq-gram distance.] A fast approximation to tree edit distance based on counting small subtree patterns.
\item[Pre-registered expectations.] Predictions written into the manuscript before the corresponding results were computed (here, before the Lean export was run).
\item[Profile$^\dagger$.] Our proposed report of proof dissimilarity: identification level, within-level distance, and a few orthogonal coordinates (area, semantics, closure size, independence), instead of a single number.
\item[Proof mining.] The extraction of explicit computational content (bounds, algorithms) from proofs, typically using proof interpretations.
\item[Proof space$^\dagger$.] The set of proofs of a theorem, relative to a library and taken up to bureaucracy, equipped with one or more of our dissimilarity measures.
\item[Proof trajectory$^\dagger$.] The sequence of intermediate lemmas, tactics and dependencies an agent accumulates while searching for a proof.
\item[Propositional truncation.] In homotopy type theory, the operation that identifies all proofs of a proposition. Lean's proof irrelevance has the same effect.
\item[Quality-diversity search.] Optimisation that seeks a collection of solutions that are both good and mutually different, rather than a single best one.
\item[Realizer; realizability.] A program that computationally witnesses a proof. Running realizers is how the kernel study extracts Herbrand witnesses.
\item[Reverse mathematics.] The programme of determining which axioms are necessary to prove a given theorem.
\item[Roles$^\dagger$: same, finer, similar, dissimilar.] The author's stipulated classification of each proof relative to the base proof; \S\ref{sec:lean-survey} proposes validating it with human ratings.
\item[Tree edit distance (TED).] The minimal number of node insertions, deletions and relabellings that transform one tree into another (Zhang--Shasha algorithm).
\item[Typeclass.] In Lean, a structure such as \texttt{Ring} or \texttt{TopologicalSpace} that is inferred automatically. Typeclass-footprint$^\dagger$ compares the classes a proof invokes.
\end{description}